\documentclass[seceq,preprint]{ptptex}
\usepackage{mathrsfs}
\usepackage[dvips]{graphicx}

  \newcommand{\sla}{\hspace{-0.5em} /}
  \newcommand{\bm}[1]{\mbox{\boldmath$#1$}}
  \newcommand{\beq}{\begin{eqnarray}}
  \newcommand{\eeq}{\end{eqnarray}}

\pubinfo{Vol.~117, No.~1, January 2007}

\notypesetlogo 

\preprintnumber[3cm]{
RBRC-635 \\ YITP-07-02}

\title{Novel Collective Excitations and the Quasi-Particle Picture
of Quarks Coupled with a Massive Boson at Finite Temperature}

\author{%
Masakiyo \textsc{Kitazawa},$^{1,2,}$\footnote{E-mail: kitazawa@quark.phy.bnl.gov}
Teiji \textsc{Kunihiro}$^{1,}$\footnote{E-mail: kunihiro@yukawa.kyoto-u.ac.jp} 
and Yukio \textsc{Nemoto}$^{3,}$\footnote{E-mail: nemoto@hken.phys.nagoya-u.ac.jp}}

\inst{$^1$Yukawa Institute for Theoretical Physics, Kyoto University, 
Kyoto 606-8502, Japan\\
$^2$RIKEN-BNL Research Center, Brookhaven National Laboratory, Upton, \\
NY 11973, USA\\
$^3$Department of Physics, Nagoya University, Naogya 464-8602, Japan}

\recdate{September 19, 2006}

\abst{%
Motivated by the observation that there may exist hadronic excitations
even in the quark-gluon plasma (QGP) phase,
we investigate how the properties of quarks,
especially within the quasi-particle
picture, are affected by the coupling with bosonic excitations
at finite temperature ($T$),  employing 
 Yukawa models with a massive scalar (pseudoscalar) and
 vector (axial-vector) boson of mass $m$.
The quark spectral function and the quasi-dispersion relations 
are calculated at one-loop order.
We find that there appears a three-peak structure in
the quark spectral function with a collective nature
when $T$ is comparable with $m$, irrespective of the type of
boson considered.  Such a multi-peak structure was first found 
in a chiral model yielding scalar composite bosons with a decay width.
We elucidate the mechanism through which
the new quark collective excitations are realized in terms of 
the  Landau damping of a quark (an antiquark) induced by
scattering with the thermally excited boson, which 
gives rise to mixing and hence a level repulsion
between a quark (antiquark) and an antiquark-hole
(quark-hole) in the thermally excited antiquark (quark) distribution.
Our results suggest that the quarks in the QGP phase 
can be described within an interesting quasi-particle picture
with a multi-peak spectral function.
Because the models employed here are rather generic,
our findings may represent a universal phenomenon for fermions 
coupled to a massive bosonic excitation with a vanishing or small width.
The relevance of these results to other fields of physics, such as
neutrino physics, is also briefly discussed.
In addition, we describe a new aspect of 
the plasmino excitation obtained in the hard-thermal  
loop approximation.
}

\begin{document}
\maketitle

\section{Introduction} \label{sec:intro}

The exploration of phase transitions in quantum chromodynamics (QCD)
and the examination of the nature of the recently discovered form of QCD 
matter under extreme conditions, such as 
high temperature ($T$) and density,
are of great interest.
 One of the central issues in this field
is revealing the physical content of the so-called 
quark-gluon plasma (QGP) phase.
Although it was widely believed for a long time that 
the QGP phase, even just after the deconfinement transition,
is composed of weakly interacting quarks and gluons,
in accordance with the asymptotic freedom of QCD,
such a view is now believed to be too naive.
In fact,
it has been being elucidated experimentally and
theoretically that the QGP phase 
near the critical point of the chiral and deconfinement
phase transitions has a non-trivial structure, because
the strong coupling nature of QCD plays a significant role.
Experiments performed at the Relativistic Heavy Ion Collider (RHIC) 
at BNL are able to probe the QGP phase 
 near the critical temperature ($T_c$)
of the deconfinement and the chiral phase transitions.
It has been argued\cite{Arsene:2004fa} that 
the experimental results at RHIC 
suggest that the created matter behaves
like a perfect fluid,
which implies that the QGP near $T_c$
is a strongly coupled system, and for this reason, it has been
called sQGP, where the ``s'' stands
for ``strongly coupled''.
This suggestion seems to be consistent with the results of
recent lattice QCD simulations\cite{Asakawa:2003re}
showing that the charmonium states  survive at higher temperature above
$T_c$ of the deconfinement transition than originally 
believed\cite{Matsui:1986dk}.
The existence of hadronic
excitations in the light quark sector in the QGP phase
has also been suggested,
on the basis of the symmetry nature of
the chiral transition,\cite{Hatsuda:1985eb}${}^,$\footnote{See also 
Ref.~\citen{DeTar:1985kx},
in which quite different reasoning is given for 
the possibility of the color-singlet excitations in 
the QGP phase.} where 
it was argued and demonstrated using a chiral effective model that if 
the chiral transition is second order or nearly second order,
there may exist soft modes of the phase transition above $T_c$
that have the same quantum numbers as the sigma meson and the pion.

Quarks and gluons are 
the basic degrees of freedom in the QGP phase, and therefore
it is fundamentally important to clarify their properties
even apart from possible hadronic excitations.
In fact, the success of the recombination (or coalescence) 
model\cite{ref:recombi}
in accounting for the 
quark-number scaling of the so-called $v_2$ parameter 
of the collective flow seen in RHIC experiments strongly suggests
that there exist quasi-particles with quark quantum numbers 
in the matter created
at RHIC.  Therefore it is imperative \cite{Kitazawa:2005mp,ref:muller-ptp}
to clarify how
the strong coupling nature of the sQGP and the quark quasi-particle
picture can be compatible. 
The present work is concerned with the properties of quarks
in a strong coupling system at finite $T$.

It is noteworthy that 
even in the high-$T$ limit, where the hard thermal loop (HTL)
approximation\cite{Braaten:1989mz} has been established,
quarks have some collective nature \cite{Bellac},
as gluons become plasmons\cite{Silin:1960, Weldon:1982aq}:
The quarks have two branches of spectra,
i.e., those of the normal quasi-quark and the plasmino\cite{Klimov:1981ka},
the latter of which does not exist at zero $T$ and density.
The origin of the plasmino excitation at
high $T$
is discussed in Ref.~\citen{Weldon:1989ys},\footnote{
The origin of a similar spectrum at zero $T$ but high density
is discussed in Ref.~\citen{Blaizot:1993bb}.
} where it is shown that 
the appearance of the plasmino at high $T$ is due to
the fact that hole states of thermally excited 
particles and anti-particles can be created.

In the vicinity of $T_c$ of the chiral phase transition,
non-perturbative effects are important, and 
one may expect, on general grounds, that
the quarks will possess
novel and complicated properties
\cite{Kitazawa:2003cs,Kitazawa:2005mp}.
In a previous work~\cite{Kitazawa:2005mp}, the present authors
explored the quark spectrum near but above $T_c$
of the chiral transition,
focusing on the effects of
the chiral soft modes\cite{Hatsuda:1985eb}.
A surprising finding in that study was that the quasi-quarks and 
quasi-antiquarks 
acquire a novel collective nature owing to the coupling of the quarks with
the soft modes, and as a result, the quark spectral function comes to have
a \textit{three-peak} structure
as $T$ approaches $T_c$\cite{Kitazawa:2005mp}.
What causes the three-peak structure in the fermion spectrum?
In reality,
the soft modes correspond to
a pronounced peak in the time-like region 
 with a width in the spectral function 
in a bosonic channel:
The peak position at the momentum $\bm{p}$
can be expressed approximately as
$\omega \simeq \pm\sqrt{ m^*_\sigma(T)^2 + |\bm{p}|^2 }$
with a $T$-dependent `mass'  $m^*_\sigma(T)$,
and as $T$ approaches $T_c$,  $m^*_\sigma(T)$
becomes smaller, together with the width of the peaks.
This implies that whenever the multi-peak structure of 
the quark spectral function appears,
the soft modes have the character of a
 well-defined elementary boson with a mass, which is
found to be comparable with $T$.\cite{Kitazawa:2005mp}${}^,$\footnote{
The bosonic excitation corresponding to a peak of the spectral function 
in the time-like region with a small width is a propagating mode with a
small damping effect. Such a mode is quite different from
those in the (color-)super-conducting transition, which are
almost diffusive, with their strength concentrated 
around the Fermi surface\cite{Kitazawa:2001ft,Kitazawa:2004cs2}.
Through a coupling with the diffusive soft mode,
the quark spectrum can form a \textit{two-peak} structure 
\cite{Kitazawa:2005pp}, which causes the pseudogap
in the density of states of quarks\cite{Kitazawa:2003cs,Kitazawa:2004cs2}.
Thus it is seen that the difference between the numbers of peaks 
in the quark spectral function
comes from the difference between the natures of the respective soft modes.
}
Thus it is seen that a smaller width of the bosonic excitation 
is favorable for the multi-peak spectral function, and 
it is feasible that a system composed of
a (massless) quark and an elementary massive boson,
as described by a Yukawa model, may exhibit
the multi-peak structure in the quark spectral function.

In this paper, we quantitatively examine
the quark properties at finite $T$ using Yukawa models
with a massive boson coupled to a massless quark.
We consider two types of bosons, i.e.,
 massive scalar (pseudoscalar) and vector (axial-vector)
 bosons with a mass $m$; note that because the quark is massless,
the scalar (vector) and pseudoscalar (axial-vector)
give the same results in the perturbation theory.
We calculate  the spectral function 
and the quasi-dispersion relation of the quark
at finite $T$ at one-loop order.
We show that a three-peak structure in the quark
spectral function is formed when $T \sim m$,
irrespective of the boson type.
We elucidate the mechanism through which
the new quark collective excitations are realized in terms of 
the Landau damping of a quark (an antiquark)
induced by scattering with thermally excited boson, which 
gives rise to mixing, and hence level repulsion,
between the quark (antiquark) and the antiquark-hole
(quark-hole) in the thermally excited
 quark (antiquark) distribution.
Such a mechanism of particle mixing 
is called ``resonant scattering''
\cite{Janko1997,Kitazawa:2005mp,Kitazawa:2005pp}.
We shall also show that the high-$T$, weak-coupling limit of 
the fermion spectrum approaches
that in the HTL approximation, and thereby
show that
the plasmino excitation obtained in the HTL
approximation can be understood as originating from  
a level repulsion between 
collective quark and anti-quark excitations.
 
Our results suggest that the quarks in the QGP phase 
can be described within an interesting quasi-particle picture with 
a multi-peak spectral function,
since there may exist bosonic excitations in the QGP phase.
Furthermore,
noting that the models employed here are rather generic,
it is natural to conjecture that 
the appearance of the novel three-peak structure in 
the fermion spectrum is a universal phenomenon for fermions 
coupled to a bosonic excitation with a vanishing or small width,
irrespective of the type of the bosonic excitation.

This paper is organized as follows.
In \S\ref{sec:spct}, the spectral function and the (quasi)-dispersion
relation of the fermion, which are frequently used in the following sections,
are summarized.
We emphasize that the quasi-dispersion relations may not describe
the physical excitations when the imaginary part of the quark
self-energy is large.
In \S\ref{sec:yukawa}, we investigate
the fermion spectrum in a Yukawa model 
 with a massive scalar boson.
We show that a three-peak structure in the fermion
spectral function is formed when  $T \sim m$.
We clarify the mechanism through which
the multi-peak structure in the spectral function is realized
in terms of  the Landau damping and the  resonant scattering.
The high-$T$, weak-coupling limit of 
the fermion spectrum is also discussed.
In \S\ref{sec:vector},
we examine the fermion spectrum in a Yukawa model with
a massive vector boson.
We find that the three-peak structure
in the fermion spectral function appears for any type of boson.
The relevance of the result to 
QGP physics is briefly discussed.
The last section is devoted to a brief summary 
and concluding remarks. 
In the appendices, some technical details of the calculations 
given in the text are presented.

\section{Generalities concerning the quark spectral function \\and  dispersion 
relations at finite temperature} \label{sec:spct}

In this section, we summarize general features of the
spectral function and
the quasi-dispersion relations of the Dirac fermion, 
which we call a `quark', at finite $T$.
Although the contents of this section may be familiar to the reader,
this section serves to introduce the notation used
in the subsequent sections.

The quark spectral function ${\cal A}( \bm{p},\omega )$ is expressed
 as\cite{Kitazawa:2004cs2}
\begin{eqnarray}
  {\cal A}( \bm{p},\omega )
  &=&-\frac1\pi {\rm Im}G^R ( \bm{p},\omega )
  \equiv -\frac1\pi \frac{ G^R( \bm{p},\omega ) 
  - \gamma^0 G^{R\dag}( \bm{p},\omega ) \gamma^0 }{2i},
\end{eqnarray}
where $G^R( \bm{p},\omega )$ is 
the retarded Green function of the quark,
\begin{eqnarray}
  G^R(\bm{p},\omega) = \frac{1}{(\omega+i\eta)\gamma^0
  -\bm{p}\cdot\mbox{\boldmath$\gamma$} + m_f
  - \Sigma^R(\bm{p},\omega)  },
\label{eq:G^R}
\end{eqnarray}
with the quark mass $m_f$ and 
the retarded self-energy $\Sigma^R(\bm{p},\omega) $.
The spectral function ${\cal A}( \bm{p},\omega )$ has 
the same matrix structure as the Green function.
Owing to the lack of Lorentz invariance at finite $T$, 
${\cal A}( \bm{p},\omega )$ generically has the Dirac structure
\begin{eqnarray}
  {\cal A}( \bm{p},\omega )
  = \rho_0( \bm{p},\omega ) \gamma^0 
  - \rho_{\rm v}( \bm{p},\omega ) \hat{\bm{p}}\cdot\bm{\gamma} +
  \rho_{\rm s}( \bm{p},\omega ),
\end{eqnarray}
where $\hat{\bm{p}} = {\bm{p}}/|{\bm{p}}|$ and
we have assumed rotational and parity invariances.
The temporal part, $\rho_0( \bm{p},\omega )$, represents
the quark number excitations of the system.

When $m_f=0$ and $ {\rm Tr} \Sigma^R = 0 $,
with ${\rm Tr}$ denoting the trace over the Dirac index,
the system possesses chiral symmetry, and
$\rho_{\rm s}( \bm{p},\omega )$ vanishes.
In this case, 
the self-energy is written as
$\Sigma^R = \Sigma_0 \gamma^0 - \Sigma_{\rm v} \bm{\gamma}\cdot\hat{\bm p}$,
and
the Green function appearing in Eq.~(\ref{eq:G^R}) can be decomposed 
using the projection operators 
$\Lambda_\pm(\bm{p}) = (1\pm\gamma^0 \hat{\bm{p}}
\cdot\bm{\gamma})/2$ as 
\begin{eqnarray}
G^R( \bm{p},\omega ) 
= G_+ (\bm{p},\omega) \Lambda_+(\bm{p})\gamma^0
+ G_- (\bm{p},\omega) \Lambda_-(\bm{p})\gamma^0,
\end{eqnarray}
with 
\begin{eqnarray}
G_\pm ( \bm{p},\omega ) 
= \frac12 {\rm Tr} [ G^R \gamma^0 \Lambda_\pm(\bm{p}) ]
= \frac1{ \omega + i\eta \mp |\bm{p}| - \Sigma_\pm ( \bm{p},\omega ) },
\end{eqnarray}
and $\Sigma_\pm ( \bm{p},\omega ) 
= (1/2) {\rm Tr} [ \Sigma^R \Lambda_\pm(\bm{p}) \gamma^0 ]
= \Sigma_0 \mp \Sigma_{\rm v}$.
The spectral function ${\cal A}( \bm{p},\omega )$ can also be written as
\begin{equation}
{\cal A}(\bm{p},\omega)
= \rho_+(\bm{p},\omega) \Lambda_+(\bm{p}) \gamma^0
+ \rho_-(\bm{p},\omega) \Lambda_-(\bm{p}) \gamma^0,
\end{equation}
where
\begin{eqnarray}
\rho_\pm ( \bm{p},\omega ) 
= -\frac{1}{\pi} {\rm Im} G_\pm
= \rho_0 ( \bm{p},\omega ) \mp  \rho_{\rm v}( \bm{p},\omega )
\label{eq:rho_pm}
\end{eqnarray}
represents the spectrum of the quark and the anti-quark excitations.
Note that $\Sigma_\pm$, $G_\pm$ and $\rho_\pm$ are scalar functions,
while $\Sigma^R$, $G^R$ and $\rho$ are matrices in the Dirac space.
In a non-interacting system,
the spectral function of the massless quark is proportional 
to the delta function:
$\rho_{\pm}^{\rm free}(\bm{p},\omega)  = \delta(\omega\mp|\bm{p}|)$.
At zero density, 
$\rho_+(\bm{p},\omega)$ and $\rho_-(\bm{p},\omega)$ have a simple
symmetric property owing to charge conjugation invariance,
\begin{equation}
  \rho_+(\bm{p},\omega) = \rho_-(\bm{p},-\omega).
  \label{eq:symm}
\end{equation}

The poles of $G_\pm (\bm{p},\omega)$, $z=z_\pm(\bm{p})$,
represent collective excitations of the quark and the antiquark sectors,
respectively.
They are obtained by solving the equations
\begin{equation}
  z_\pm(\bm{p}) \mp |\bm{p}|-
  \Sigma_\pm(\bm{p},z_\pm(\bm{p})) = 0
  \label{eq:pole}
\end{equation}
in the complex energy plane.
The form of the retarded self-energy guarantees that the solutions $z_\pm$
lie in the lower-half plane, owing to causality.
Equation~(\ref{eq:pole}) may have several solutions
for fixed momentum $|\bm{p}|$, and $\rho_\pm (\bm{p},\omega)$
may have a multi-peak structure associated with these poles.

For convenience, we also define 
the quasi-dispersion relations 
$\omega=\omega_\pm(\bm{p})$
as  the zero of the {\it real part} of the inverse Green functions:
\begin{equation}
  \omega_\pm(\bm{p}) \mp |\bm{p}| =
  \textrm{Re}\Sigma_\pm(\bm{p},\omega_\pm(\bm{p})).
  \label{eq:disp}
\end{equation}
Solving Eq.~(\ref{eq:disp}) is much easier than solving
Eq.~(\ref{eq:pole}) in the complex plane.
The functions $\omega_\pm(\bm{p})$ are real numbers,
and Eq.~(\ref{eq:disp}) may have several solutions for fixed $\bm{p}$.
When the condition
$\textrm{Im}\Sigma_\pm(\bm{p},\omega_\pm(\bm{p}) ) \ll 
\omega_\pm(\bm{p})$ is satisfied, and hence 
the quasi-particle picture in the 
Landau sense is valid,
the difference between Re$z_\pm$ and $\omega_\pm$ is small.
Then $\omega_\pm(\bm{p})$ can be regarded
as an approximation of the excitation energy of the quasi-particles and
can be used to estimate positions of the peaks in $\rho_\pm$.\footnote{
In this sense, when this condition is satisfied,
the peaks of the spectral function and the quasi-dispersion
relation, which are gauge-dependent in the gauge theories, accurately reflect 
the position of the gauge-independent poles of the Green function.}
We often use the quasi-dispersion relations in this paper.

\section{Quark spectrum in a Yukawa model with a massive
scalar (pseudoscalar)  boson} 
\label{sec:yukawa}

In this section, we investigate the spectrum of a massless quark
$\psi$ in a Yukawa model with a massive scalar (pseudoscalar)
 boson $\phi$ at finite $T$.
A detailed analysis of a {\em massive quark}
coupled with a {\em massless boson}
 at vanishing momentum is given in 
Ref.~\citen{Baym:1992eu}.
As we find below, the finite boson mass gives rise to
unexpectedly interesting complications in
the quark spectrum, which were not seen in Ref.~\citen{Baym:1992eu}.

We start from the following Lagrangian composed of a quark
and a scalar boson:
\begin{equation}
  \mathscr{L} = \bar{\psi} (i \partial\sla - g \phi) \psi
  + \frac12 \left( \partial_\mu \phi \partial^\mu \phi - m^2 \phi^2 \right).
\label{eq:lag}
\end{equation}
Here, $g$ is the coupling constant and $m$ is the boson mass.
We note that if the boson is a pseudoscalar,
$g$ should be replaced with $i\gamma_5g$. 
As we shall see, however, this replacement does not lead to
any difference in the results for the quark spectrum.

\begin{figure}[t]
\begin{center}
\includegraphics[width=140pt]{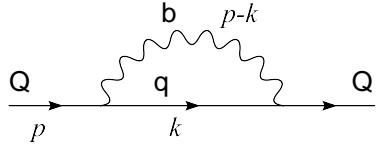}
\caption{
The quark self-energy at one-loop order.
}
\label{fig:loop}
\end{center}
\end{figure}

\subsection{Quark self-energy}
\label{subsec:selfe}

\subsubsection{Formulation}

The quark self-energy in the imaginary time formalism
at one-loop order, shown in Fig.~\ref{fig:loop}, is given by
\begin{eqnarray}
\tilde\Sigma ( \bm{p},i\omega_m )
= -g^2 T \sum_n \int \frac{ d^3\bm{k} }{ (2\pi)^3 }
{\cal G}_0 ( \bm{k},i\omega_n )
{\cal D}( \bm{p}-\bm{k} ,i\omega_m-i\omega_n ),
\label{eq:tildeSigma}
\end{eqnarray}
where 
${\cal G}_0 ( \bm{k},i\omega_n ) 
= [ i\omega_n \gamma^0 - \bm{k}\cdot\bm{\gamma} ]^{-1}$
and 
${\cal D} ( \bm{k},i\nu_n ) = [ (i\nu_n)^2 - \bm{k}^2 - m^2 ]^{-1}$
are the Matsubara Green functions for the free quark and scalar boson,
respectively, and
$\omega_n = (2n+1)\pi T$ and $\nu_n = 2n\pi T$ are the
Matsubara frequencies for the fermion and boson, respectively.

For the coupling with a pseudoscalar boson, a factor of
$i\gamma_5$ appears on both sides of ${\cal G}_0$ in 
Eq.~(\ref{eq:tildeSigma}).
However, these factors cancel out because they anti-commute with
${\cal G}_0$, and thus the self-energy takes the same form 
as Eq.~(\ref{eq:tildeSigma}).
Therefore, the following results and arguments in this section hold 
also for coupling with a pseudoscalar boson,
as mentioned above.

In accordance with the standard procedure in the imaginary time
formalism, we carry out the summation over the Matsubara modes $n$ and the
analytic continuation $i\omega_m\to\omega+i\eta$
(see Appendix~\ref{app:sigma} for details). 
The result is given by
\begin{align}
  \Sigma^R(\bm{p},\omega) &= g^2 \int \frac{ d^3\bm{k} }{(2\pi)^3}
  \frac{1}{2E_b}\bigg\{ \Lambda_+(\bm{k})
  \frac{ 1+n(E_b)-f(E_f) }{ \omega-E_f-E_b+i\eta } +
  \Lambda_-(\bm{k})
  \frac{ n(E_b)+f(E_f) }{ \omega+E_f-E_b+i\eta}
  \notag \\
  &\   +\Lambda_+(\bm{k})
  \frac{ n(E_b)+f(E_f) }{ \omega-E_f+E_b+i\eta }
  +\Lambda_-(\bm{k})
  \frac{ 1+n(E_b)-f(E_f) }{ \omega+E_f+E_b+i\eta } 
  \bigg\} \gamma^0,
  \label{eq:sigmar}
\end{align}
where $E_f=|\bm{k}|$, $E_b=\sqrt{(\bm{p}-\bm{k})^2+m^2}$,
and 
$n(E)=[\exp(E/T)-1]^{-1}$ and $f(E)=[\exp(E/T)+1]^{-1}$
are the Bose-Einstein and Fermi-Dirac distribution functions,
respectively.

The self-energy given in Eq.~(\ref{eq:sigmar}) has an ultraviolet divergence,
which comes from the $T$-independent part, 
$ \Sigma^R(\bm{p},\omega)_{T=0}
\equiv \lim_{ T\to 0 }\Sigma^R(\bm{p},\omega)$.
To eliminate this divergence,
we renormalize this part by imposing the on-shell 
renormalization condition.
(See Appendix~\ref{app:T=0} for details of the renormalization
procedure.)
The renormalized self-energy is given by
\begin{eqnarray}
\Sigma^R(\bm{p},\omega)_{T=0} =
\frac{g^2 p\sla}{32\pi^2} \bigg\{
\bigg( 1-\frac{m^2}{P^2} \bigg)^2 \ln \frac{P^2-m^2}{m^2}
-\frac{3}{2} + \frac{m^2}{P^2}  \bigg\},
\label{eq:T=0part}
\end{eqnarray}
with $p_\mu = ( \omega , \bm{p} )$ and $P^2 = p_\mu p^\mu $.
Note that there is no infrared divergence in Eq.~(\ref{eq:sigmar})
as long as the scalar boson is massive.
Since the $T$-dependent $(T\ne0)$ part,
$ \Sigma^R(\bm{p},\omega)_{T\ne0}
\equiv \Sigma^R(\bm{p},\omega) - \Sigma^R(\bm{p},\omega)_{T=0}$,
has no divergences, 
we can evaluate it without regularization
(see Appendix~\ref{app:sigma}).
For the numerical calculations of the $T\ne0$ part, 
it is convenient to first derive the imaginary part,
${\rm Im}\Sigma^R (\bm{p},\omega)_{T\ne0}$,
and then determine the real part using the dispersion
(Kramers-Kronig) relation
\begin{eqnarray}
{\rm Re}\Sigma^R( \bm{p},\omega )_{T\ne0}
= - \frac1\pi {\rm P}\int_{-\infty}^\infty d\omega' 
\frac{ {\rm Im}\Sigma^R( \bm{p},\omega' )_{T\ne0} }{ \omega - \omega' },
\label{eq:Kramers-Kronig}
\end{eqnarray}
where the symbol P denotes the principal value.

\begin{figure}[t]
\begin{center}
\includegraphics[width=100pt]{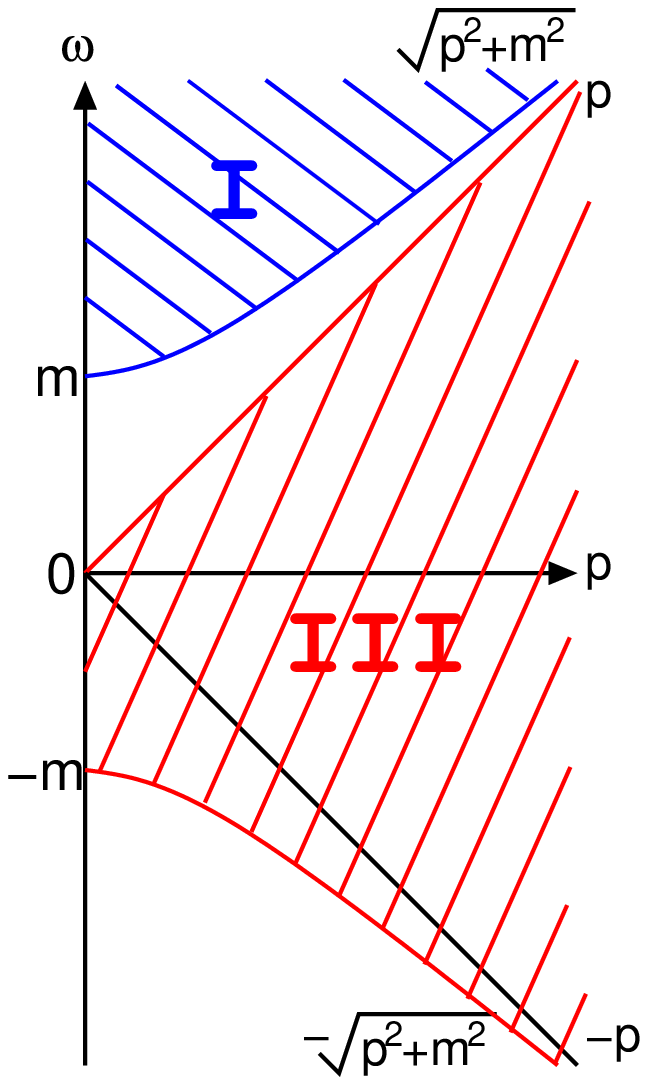}
\includegraphics[width=100pt]{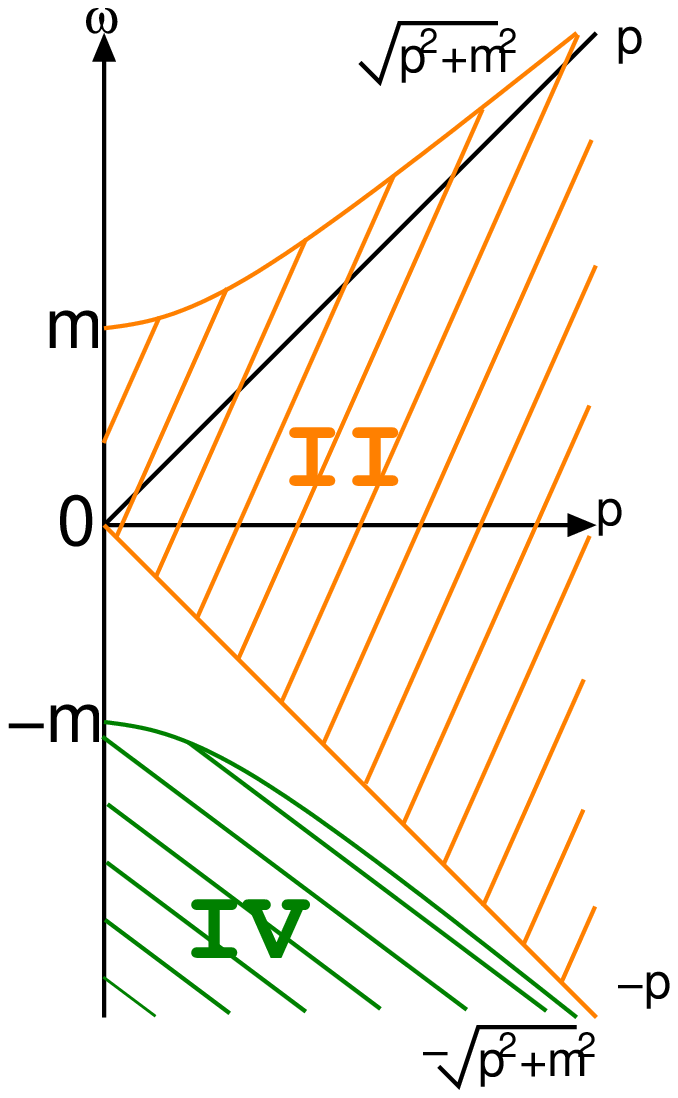}
\caption{
The supports of the terms (I)-(IV) in Eq.~(\ref{eq:imsig}).
}   
\label{fig:imsig}
\end{center}
\end{figure}

The imaginary part of the self-energy,
${\rm Im}\Sigma^R (\bm{p},\omega)$, is given by
\begin{align}
  \textrm{Im}\Sigma^R(\bm{p},\omega) &= -\pi g^2
  \int \frac{d^3k}{(2\pi)^3} \frac{1}{2E_b} \notag \\
  &\ \times \bigg\{ \Lambda_+(\bm{k}) 
  (1+n-f) \delta(\omega-E_f-E_b)
  \qquad \textrm{(I)} \notag \\
  &\ +\Lambda_-(\bm{k}) (n+f)
  \delta(\omega+E_f-E_b) \hspace{1.725cm} \textrm{(II)} \notag \\
  &\ +\Lambda_+(\bm{k}) (n+f)
  \delta(\omega-E_f+E_b) \hspace{1.725cm} \textrm{(III)}  \notag \\
  &\ +\Lambda_-(\bm{k}) (1+n-f)
  \delta(\omega+E_f+E_b) \bigg\} \gamma^0 \quad \textrm{(IV)},
  \label{eq:imsig}
\end{align}
with $f=f(E_f)$ and $n=n(E_b)$.
Each term in Eq.~(\ref{eq:imsig}), denoted by (I)-(IV), describes a definite 
decay process and has support in the region shown in 
Fig.~\ref{fig:imsig}.
The details of the physical meaning of each term are discussed
in the following.

\begin{figure}[t]
\begin{center}
\includegraphics[width=200pt]{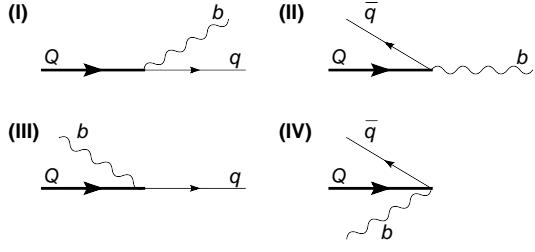}
\caption{
The kinetic processes contained in Im$\Sigma_+(\bm{p},\omega)$.
(The corresponding inverse ones are not shown.)
The thick solid line labeled $Q$ represents
the quasi-quark,
the thin solid lines labeled $q$ on-shell free quarks, and 
the wavy lines labeled $b$ the on-shell 
scalar boson. The incident on-shell particles are thermally excited 
particles.}   
\label{fig:feynykw}
\end{center}
\end{figure}

Let us denote the external quark as $Q$ and the internal quark as $q$,
as shown in Fig. \ref{fig:loop}.
The former is the quasi-quark interacting with scalar bosons,
while the latter is a free quark in the approximation used here.
The physical meaning of each term in Eq.~(\ref{eq:imsig}) is
more transparent when
Eq.~(\ref{eq:imsig}) is rewritten in terms of the transition
probabilities \cite{Weldon:1983jn}.
We can express $\Sigma_+$ defined in \S\ref{sec:spct} as
\begin{equation}
  \Sigma_+(\bm{p},\omega) = \bar{u}(p)\Sigma^R(\bm{p},\omega)u(p),
  \label{eq:sigma+}
\end{equation}
where $u(p)$ is a free quark spinor with the effective mass 
parameter $\sqrt{P^2}$,
which satisfies
$(p\sla-\sqrt{P^2})u(p)=0$ with the normalization $\bar{u}(p)u(p)=2\sqrt{P^2}$.
Then, we have
\begin{align}
  \textrm{Im}\Sigma_+(\bm{p},\omega) &=
  -\frac{1}{2} \int \frac{d^3k_f}{(2\pi)^3} \frac{d^3k_b}{(2\pi)^3}
  \frac{1}{2E_f 2E_b} \sum_{s} (2\pi)^4 \notag \\
  &\ \times \bigg\{
  [\delta^4(p-k_f-k_b) |M(Q\to q,b)|^2 [(1-f)(1-n)+fn]
  \hspace{1.5cm}\textrm{(I)} \notag \\
  &\ +
  [\delta^4(p+k_f-k_b) |M(Q,\bar{q}\to b)|^2 
  [f(1+n)+n(1-f)] \hspace{1.725cm} \textrm{(II)} \notag \\
  &\ +
  [\delta^4(p-k_f+k_b) |M(Q,b\to q)|^2 [n(1-f)+f(1+n)]
  \hspace{1.725cm}\textrm{(III)} \notag \\
  &\ +
  [\delta^4(p+k_f+k_b) |M(Q,\bar{q},b\to 0)|^2
   [fn+(1-f)(1-n)] \bigg\}  \hspace{1.1cm} \textrm{(IV)},
  \label{eq:impiu}
\end{align}
where $s$ is the spin of the quark $q$ and
$k_{f}=(E_{f},\bm{k}),
k_b=(E_b,\bm{p}-\bm{k})$.
The amplitudes $M$ are given by
\begin{eqnarray}
  M(Q\to q,b) &=& M(Q,b\to q) = g\bar{u}(p)u(p), \nonumber \\
  M(Q,\bar{q}\to b) &=& M(Q,\bar{q},b\to 0) = g\bar{v}(p)u(p),
\end{eqnarray}
where $v(p)$ is defined through
$(p\sla+\sqrt{P^2})v(p)=0$, with the normalization 
$\bar{v}(p)v(p)=-2\sqrt{P^2}$.
The first term in Eq.~(\ref{eq:impiu}) with the statistical factor 
$(1-f)(1+n)$ describes a decay process $Q\to q+b$, and that with
the factor $fn$ describes the inverse one, $q+b\to Q$.
The second term with the factor $f(1+n)$ describes $Q+\bar{q}\to b$, 
and that with the factor $n(1-f)$ describes $b\to Q+\bar{q}$.
The third and the fourth terms are similar.
These processes are schematically depicted in Fig. \ref{fig:feynykw}.
The terms (II) and (III), which involve a thermally excited particle
as the incident particle, vanish as $T\to0$ and are known as the 
Landau damping.
They play an important role in the structure of the quark
spectral function at finite $T$, as we see below.

Similarly, it can be shown that the quantity
$\Sigma_-(\bm{p},\omega) = \bar{v}(p)\Sigma^R(\bm{p},\omega)v(p)$
contains amplitudes for the processes
$M(\bar{Q}\to \bar{q},b)$,
$M(\bar{Q},q\to b)$,
$M(\bar{Q},b\to \bar{q})$, and
$M(\bar{Q},q,b\to 0)$.

After some manipulations as given in Appendix~\ref{app:sigma},
we arrive at the following result
\begin{eqnarray}
{\rm Im}\Sigma_\pm ( \bm{p},\omega )
&=&
 - \frac{ g^2 }{ 32\pi \bm{p}^2 }
\int_{E_f^+}^{E_f^-} dE_f [2|\bm{p}| E_f \mp
  (\bm{p}^2 + m^2 - \omega^2 + 2\omega E_f )] \notag \\
& & \qquad \times  [ 1 + n(\omega-E_f ) - f( E_f ) ] \notag \\
& &- \frac{ g^2 }{ 32\pi \bm{p}^2 } 
[(|\bm{p}|\mp \omega)( \pi^2 T^2 \pm \omega^2 ) \pm \omega m^2]
\theta( \bm{p}^2 - \omega^2 ),
\label{eq:ImSig0Y_simple}
\end{eqnarray}
with $E_f^\pm = ( \omega^2 - \bm{p}^2 - m^2 ) / 2( \omega \pm |\bm{p}| )$.
For vanishing momentum, we have ${\rm Im}\Sigma_+ = {\rm Im}\Sigma_-$,
and ${\rm Im}\Sigma^R$ is simply given by
\begin{align}
{\rm Im}\Sigma^R ( \bm{p}=0,\omega )
&= -\gamma^0 \frac{g^2}{32\pi}
\frac{ (\omega^2-m^2 )^2 }{ \omega^3 }
\left[ n\left( \omega_b \right)
+ f\left( \omega_{q} \right) \right],
\label{eq:ImSigY_p0_simple}
\end{align}
with
\begin{equation}
\omega_b = \frac{\omega^2 + m^2}{2\omega}, \qquad
\omega_q = -\frac{\omega^2 - m^2}{2\omega}.
\label{eq:omega_bq}
\end{equation}

\begin{figure}[t]
\begin{center}
\includegraphics[width=0.66\textwidth]{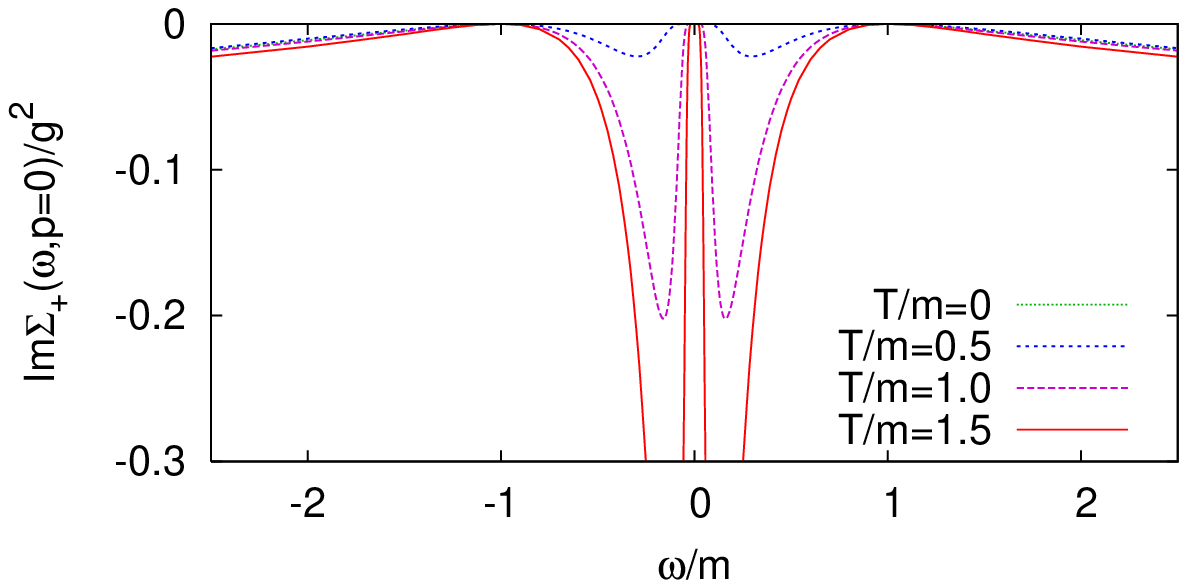}
\includegraphics[width=0.66\textwidth]{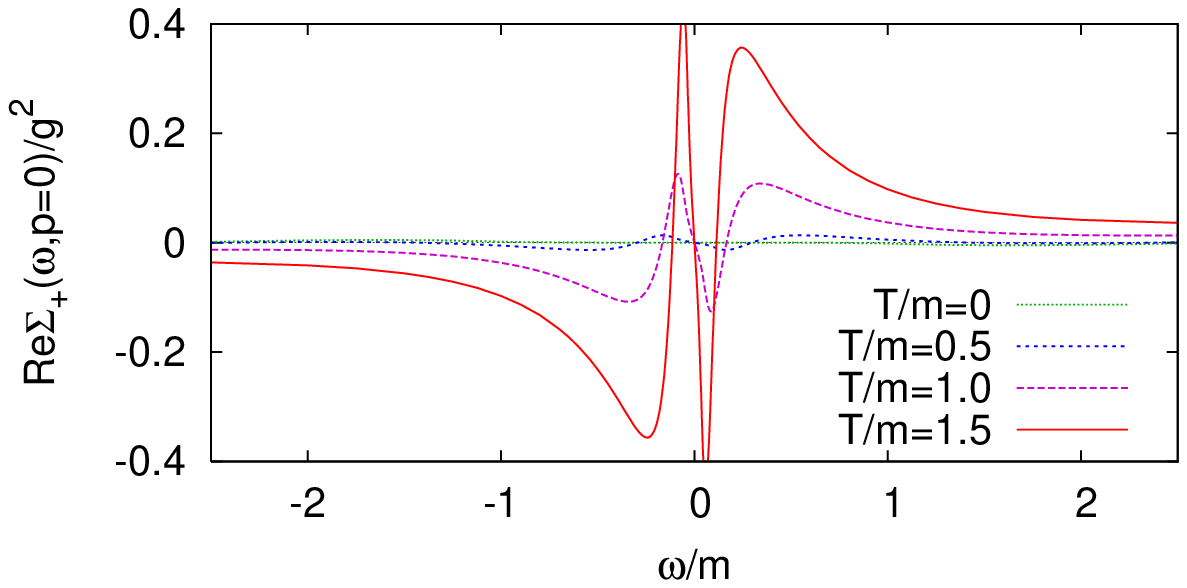}
\caption{
The imaginary and real parts of the self-energy 
$\Sigma_+ (\bm{p},\omega)$ at zero momentum
and several temperatures.
}
\label{fig:ImSigma_p0}
\end{center}
\end{figure}

\subsubsection{Numerical results}

In the upper panel of Fig.~\ref{fig:ImSigma_p0},
we plot Im$\Sigma_+ (\bm{p},\omega)/g^2$, which is independent of
$g$, for vanishing momentum and several temperatures.
We see that Im$\Sigma_+ (\bm{0},\omega)/g^2$ vanishes 
at $\omega=0,\pm m$ for all $T$.
These values of $\omega$ are located in the boundaries of the regions
(I)--(IV) shown in Fig.~\ref{fig:imsig}.
The vanishing decay rate at $\omega=\pm m$ is due to
the suppression of the phase space;
energy-momentum conservation requires
vanishing momentum of the on-shell (anti)quark
for each process at $\omega=\pm m$.
Near $\omega=0$, by contrast, 
the distribution functions in Eq.~(\ref{eq:ImSigY_p0_simple}) 
suppress the decay rate,
because the on-shell energies of the boson and the quark,
$|\omega_b|$ and $|\omega_q|$, go to infinity as $\omega\to 0$.

From Fig.~\ref{fig:ImSigma_p0}, we see that 
Im$\Sigma_+ (\bm{p},\omega)$ takes finite values 
only for $|\omega|/m>1$ at $T=0$;
the corresponding decay processes 
are represented by (I) and (IV) in Fig.~\ref{fig:feynykw}.
At finite $T$, the Landau damping has a significant effect,
and Im$\Sigma_+ (\bm{p},\omega)$ has support in the regions
(II) and (III) in Fig.~\ref{fig:imsig}.
The height of the two peaks in these regions increase rapidly as 
$T$ increases, 
and for $T\simeq m$, the Landau damping comes to dominate
the decay rates, owing to the processes (I) and (IV)
in the energy range shown in Fig.~\ref{fig:ImSigma_p0}.

Using the dispersion relation given in Eq.~(\ref{eq:Kramers-Kronig}),
we can understand the qualitative behavior of 
${\rm Re}\Sigma_+ ( \bm{p},\omega )$ from that of 
the imaginary part:
If there is a sharp peak in the imaginary part, 
the real part has a steep increase at the same energy.
In order to see this relation,
we plot ${\rm Re}\Sigma_+ ( \bm{p},\omega )/g^2$ for $\bm{p}=0$
in the lower panel of Fig.~\ref{fig:ImSigma_p0}.
It is seen that 
there appears oscillating behavior around $\omega=0$, and
the amplitude of this oscillation grows rapidly 
as $T$ increases,
along with the  growth of the peaks 
in ${\rm Im}\Sigma_+( \bm{p},\omega )$.

The other important property of $\Sigma_+ ( \bm{p},\omega )$
shown in Fig.~\ref{fig:ImSigma_p0} is that both 
${\rm Re}\Sigma_+ ( \bm{0},0 )$ and 
${\rm Im}\Sigma_+ ( \bm{0},0 )$
vanish for all $T$.
The fact that $\Sigma^R ( \bm{0},0 )$ vanishes implies the existence of a
pole of the quark propagator at the origin (see Eq.~(\ref{eq:pole})).
(The reason that ${\rm Im}\Sigma_+( \bm{0},0 )$ vanishes
is explained above.)
The vanishing of the real part can be easily understood from
the dispersion relation and the fact that 
${\rm Im}\Sigma_+(\bm{0},\omega)$ is an even function of $\omega$.
Thus we find that
Re$\Sigma_+ ( \bm{0},\omega )$ is an odd function of $\omega$.

\begin{figure}[t]
\begin{center}
\includegraphics[width=0.7\textwidth]{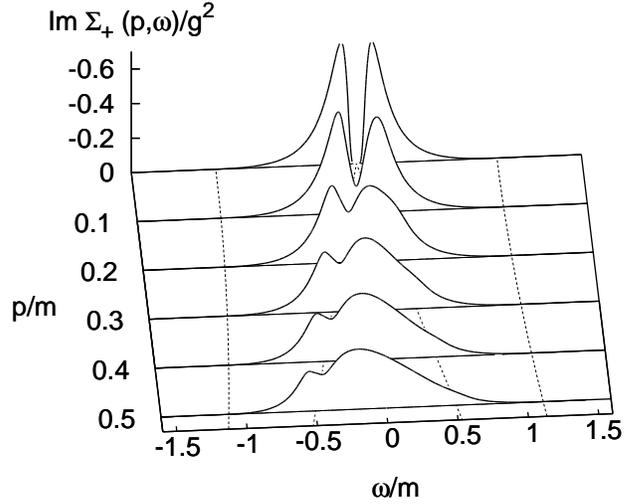}
\caption{
The imaginary part of the self-energy $|{\rm Im}\Sigma_+ (\bm{p},\omega)|$ 
for $T/m=1.5$.
The dashed curves in the $\omega$-$\bm{p}$ plane 
denote $ \omega=\pm\sqrt{ |\bm{p}|^2 + m^2 } $
and $ \omega = \pm\bm{p} $.
}
\label{fig:ImSigma3d}
\end{center}
\end{figure}

In Fig.~\ref{fig:ImSigma3d}, 
we display $|{\rm Im}\Sigma_+ (\bm{p},\omega)|/g^2$
at finite momenta and energies for $T/m=1.5$.
The two separated peaks at $\bm{p}=\bm{0}$, coming from the Landau damping,
tend to overlap, and are smeared as $\bm{p}$ becomes large.

\subsubsection{Asymptotic behavior in the high temperature,
weak coupling limit}

Finally, let us consider the asymptotic behavior of $\Sigma^R(\bm{p},\omega)$
in the high $T$, weak coupling limit, i.e.
$gT\to\infty$ and $T/m \to \infty$, with $g^2T\to0$.

In this limit, the leading contribution to the imaginary part,
${\rm Im}\Sigma^R(\bm{p},\omega)$, comes from
the terms which explicitly include the factor $T^2$ in 
Eq.~(\ref{eq:ImSig0Y_simple}).
Thus, the asymptotic form of ${\rm Im}\Sigma^R( \bm{p},\omega )$ 
is given by
\begin{eqnarray}
{\rm Im}\Sigma^R ( \bm{p},\omega ) |_{ T\to\infty }
=- \frac{ \pi g^2 T^2 }{ 32 } 
\left( \frac{ \gamma^0 }{ |\bm{p}| } 
- \frac{ \omega \hat{\bm p}\cdot\bm{\gamma} }{\bm{p}^2} \right)
\theta( \bm{p}^2 - \omega^2 ).
\label{eq:ImSig_highT}
\end{eqnarray}
The real part in this limit is easily calculated using
Eq.~(\ref{eq:ImSig_highT}) 
and the dispersion relation given in Eq.~(\ref{eq:Kramers-Kronig}).
We then obtain the following asymptotic form of 
$\Sigma^R(\bm{p},\omega)$:
\begin{equation}
  \Sigma^R(\bm{p},\omega) |_{ T\to\infty }
  = m_T^2 \frac{\gamma^0}{|\bm{p}|}Q(\bm{p},\omega) +
  m_T^2 \frac{\hat{\bm{p}}\cdot\bm{\gamma}}{|\bm{p}|}
  \bigg( 1-\frac{\omega}{|\bm{p}|} Q(\bm{p},\omega) \bigg),
  \label{eq:sigmahtl}
\end{equation}
with the thermal mass $m_T = gT/4$ and
\begin{equation}
  Q(\bm{p},\omega) = \frac{1}{2}\ln \bigg| \frac{\omega+|\bm{p}|}
  {\omega-|\bm{p}|}\bigg| - i \frac{\pi}{2}\theta(\bm{p}^2-\omega^2).
  \label{eq:Q}
\end{equation}
Equation~(\ref{eq:sigmahtl}) is the same as
the fermion self-energy of QED and QCD in the HTL approximation
\cite{Bellac}, 
apart from the definition of the thermal mass.
Therefore, the quark spectrum in the Yukawa model 
in the limit $gT\to\infty$ ($T/m\to \infty$) with $g\to0$ 
is equivalent to the well-known spectrum
in the gauge theories \cite{Weldon:1982aq}.

We note that the analytic structure of the self-energy
is qualitatively different from that at intermediate $T$, i.e. 
for $T\simeq m$;
there is no pole of the propagator at the origin in this limit,
because Re$\Sigma_+(\bm{0},\omega\to0)$ behaves as
$\sim 1/\omega$ and hence never vanishes,
while Im$\Sigma_+(\bm{0},\omega\to0)$ vanishes.
This change in the self-energy is reflected in that of 
the peak structure
of the spectral function, as discussed in the next subsection.

\subsection{Quark spectral function at various temperatures}

In this subsection,
we examine how the quasi-particle picture of the quark changes
at finite $T$ by studying the quark spectral function
$\rho_\pm( \bm{p},\omega )$
and the quasi-dispersion relation $\omega_\pm(\bm{p})$.
We fix the coupling constant as $g=1$ throughout this subsection
to elucidate the $T$ dependence of the spectrum.
The $g$ dependence will be discussed in a subsequent subsection.
Because the only dimensional parameter in Eq.~(\ref{eq:lag})
is the boson mass $m$, we scale all the dimensional
parameters $T,\omega$ and $\bm{p}$ by $m$.

\subsubsection{Quark spectral function at zero 
temperature and the hard thermal loop
approximation}

Before presenting the numerical results,
we briefly review the quark spectrum at $T=0$ and 
high $T$ limit in our model.
The analytic form of the quark self-energy at $T=0$ is
given in Eq.~(\ref{eq:T=0part}).
In this case, the poles of the quark propagator 
are on the light cone, $\omega=\pm|\bm{p}|$, 
because we have employed the on-shell renormalization condition 
to derive the self-energy appearing in Eq.~(\ref{eq:T=0part}).
The quark spectral functions $\rho_\pm(\bm{p},\omega)$ 
are then given by
\begin{eqnarray}
  \rho_\pm( \bm{p},\omega ) 
  = Z(0)\cdot \delta( \omega \mp |\bm{p}| ) 
  + \rho_\pm^{\rm (cont)}( \bm{p}, \omega ),
\label{eq:freerho}
\end{eqnarray}
 where the $T$-dependent residue $Z(T)$ is exactly unity
 at $T=0$ under the on-shell renormalization condition, 
and the continuum part $\rho_\pm^{\rm (cont)}$ 
takes finite values for $|\omega|>\sqrt{ \bm{p}^2 + m^2 }$.
\footnote{
Some readers may be uneasy about the violation of
the sum rule for the spectral function,
$
  \int_{-\infty}^\infty d\omega \rho_\pm(\bm{p},\omega) =1.
$
It should be noted here that
in relativistic field theory, this sum rule 
does not necessarily hold after the renormalization\cite{Bellac}.}

In the high $T$, weak coupling limit, 
the quark spectrum approaches that calculated 
in the HTL approximation in the gauge theories
with the thermal mass $m_T=gT/4$, as does the quark self-energy 
(see the previous subsection).
In this limit, both $\rho_+( \bm{p},\omega )$ and $\rho_-( \bm{p},\omega )$
have two delta-functions corresponding to the normal quasi-particle and 
plasmino excitations, in addition to the continuum part in the
space-like region\cite{Bellac,Weldon:1989ys} 
(see Eqs.~(\ref{eq:sigmahtl}) and (\ref{eq:Q})).
In the high $T$ limit with fixed $g$,
these two peaks have widths of the order of $gm_T$\cite{Bellac}.

\begin{figure}[t]
\begin{center}
\begin{tabular}{cc}
\hspace{-1.8cm} $T/m=0.4$ & \hspace{-2.3cm} $T/m=0.8$ \\
\hspace{-.7cm}
\includegraphics[width=220pt]{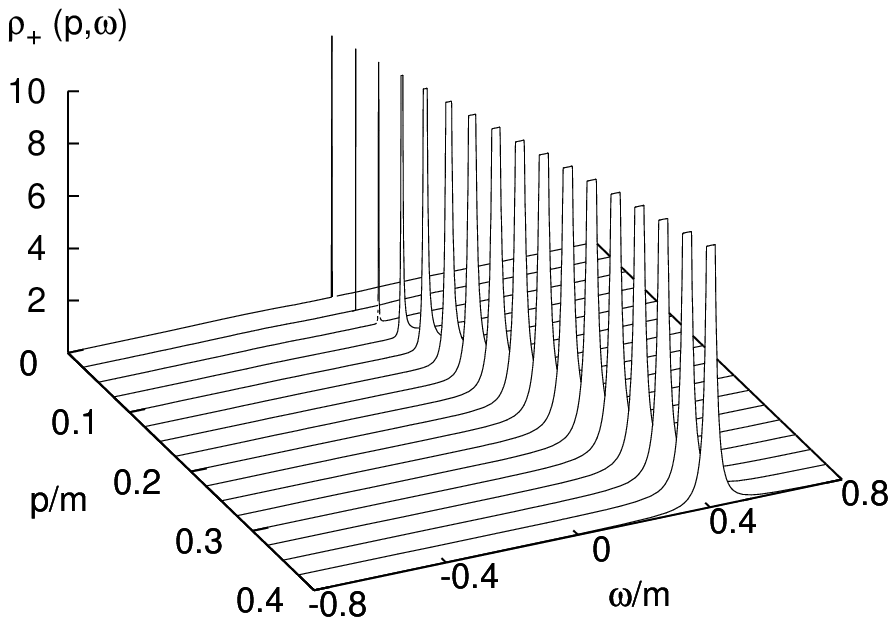} & \hspace{-1.3cm}
\includegraphics[width=220pt]{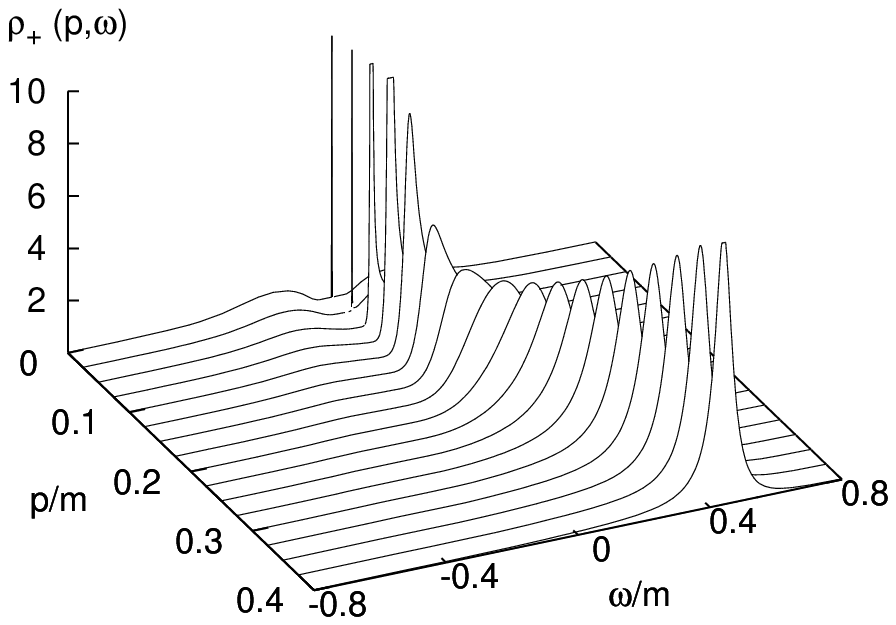} 
\end{tabular}
\caption{
The quark spectral function $\rho_+(\bm{p},\omega)$ for 
$T/m=0.4$ and $T/m=0.8$.
}
\label{fig:spc1}
\end{center}
\end{figure}

\begin{figure}
\begin{center}
\includegraphics[width=280pt]{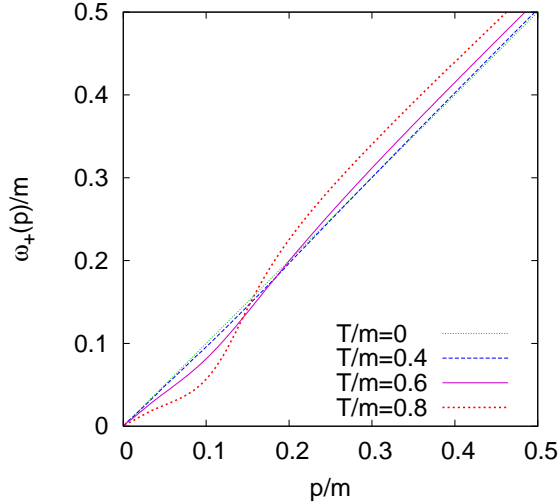}
\caption{
The quasi-dispersion relation
$\omega_+(\bm{p})$ for several values of $T$.
}
\label{fig:qdr1}
\end{center}
\end{figure}

\subsubsection{Quark spectral function at low temperature}

\begin{figure}[t]
\begin{center}
\begin{tabular}{cc}
$T/m=0.8$ & $T/m=1.4$ \\
\includegraphics[width=0.49\textwidth]{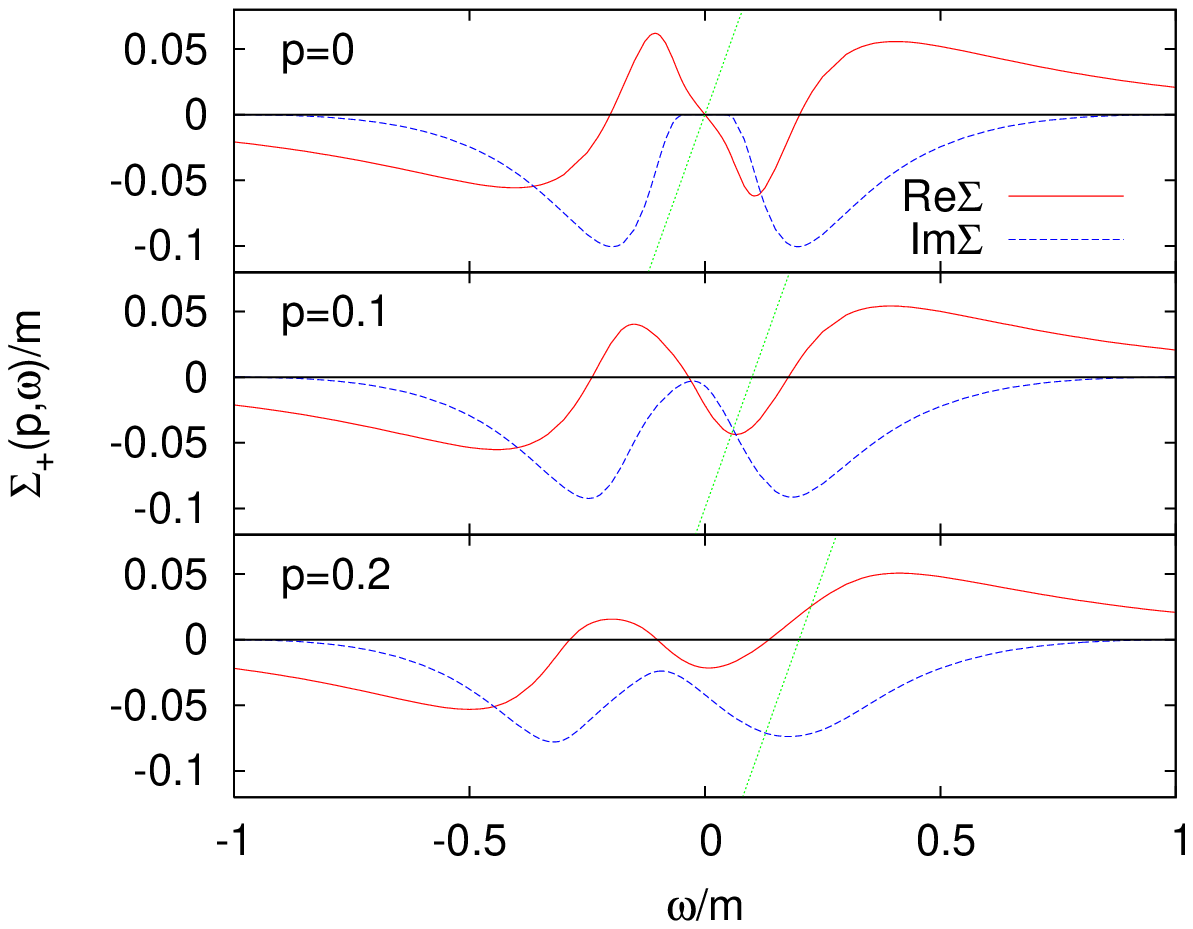} &
\includegraphics[width=0.49\textwidth]{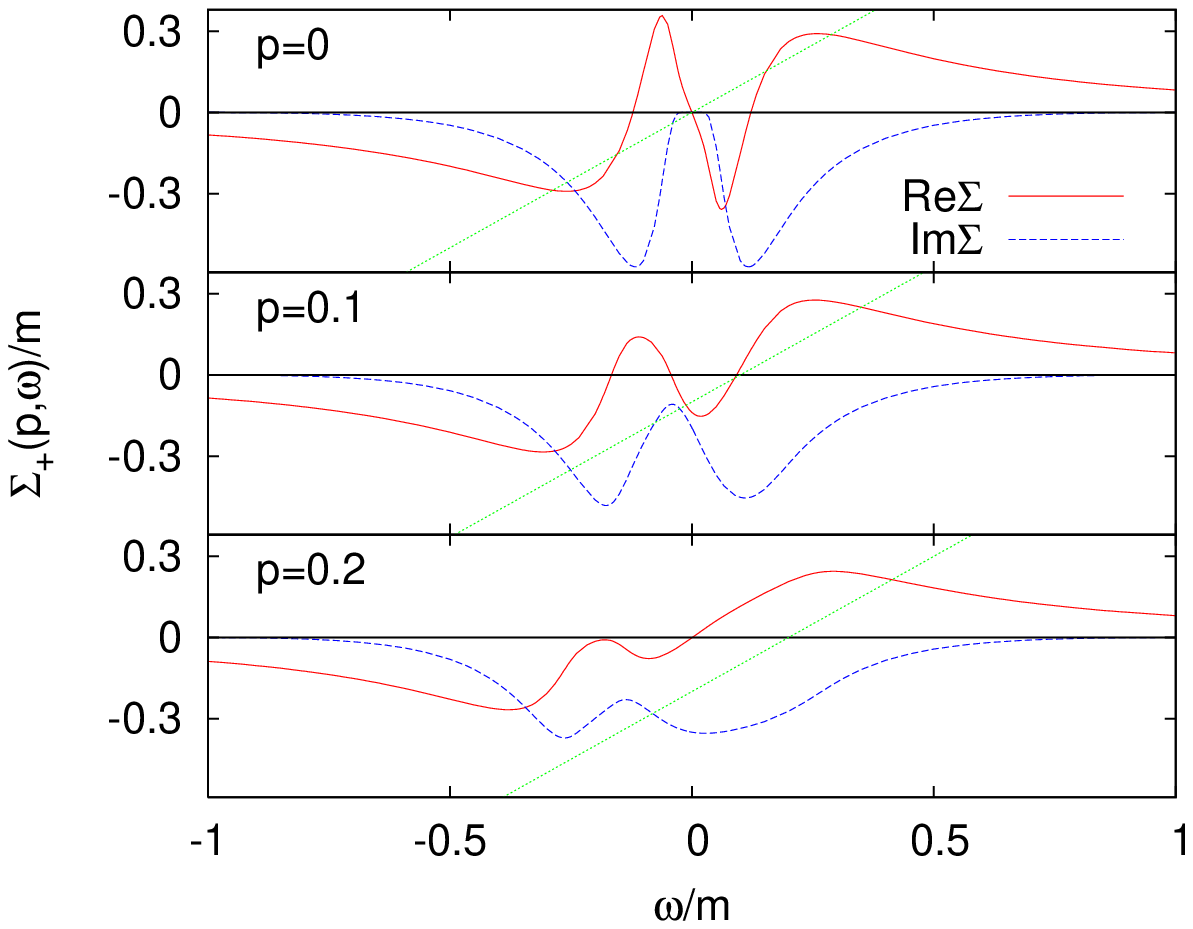}
\end{tabular}
\caption{
The quark self-energy $\Sigma_+(\bm{p},\omega)$ for several
momenta 
with $T/m=0.8$ and $1.4$. The dotted line represents the l.h.s.
of Eq.~(\ref{eq:disp}), i.e., $\omega-|\bm{p}|$.
}
\label{fig:self}
\end{center}
\end{figure}

In Figs. \ref{fig:spc1} and \ref{fig:qdr1},
we plot the spectral function $\rho_+(\bm{p},\omega)$ 
and the quasi-dispersion relation $\omega_+(\bm{p})$
of the quark part for relatively low temperatures, i.e., $T<m$.
The anti-quark parts $\rho_-$ and $\omega_-(\bm{p})$
can be determined from the symmetric properties given in Eq.~(\ref{eq:symm})
and the relation $\omega_-(\bm{p}) = -\omega_+(\bm{p})$.
At $T/m=0.4$, the spectral function $\rho_+( \bm{p},\omega )$ and
$\omega_+(\bm{p})$ are almost the same as those at $T=0$;
$\rho_+( \bm{p},\omega )$ has sharp quasi-particle peaks 
around $\omega=|\bm{p}|$, and $\omega_+(\bm{p})\simeq |\bm{p}|$.
At $T/m=0.8$, 
the quasi-particle peaks are clearly deformed near $|\bm{p}|/m=0.15$.
The quasi-dispersion relation also deviates
from the free one near $|\bm{p}|/m=0.15$.
We note that this relation exhibits an unphysical acausal behavior;
the imaginary part of
the self-energy is large in this momentum region, 
as shown in Fig.~\ref{fig:self}, and thus
the quasi-dispersion relation becomes unphysical in this region.

To understand the behavior of $\rho_+( \bm{p},\omega )$ and
$\omega_+(\bm{p})$ depicted in Figs.~\ref{fig:spc1} and \ref{fig:qdr1},
we show the real and imaginary parts of $\Sigma_+(\bm{p},\omega)$
at $T/m=0.8$ and several momenta in the left panel of
Fig.~\ref{fig:self}.
We see that there are two peaks 
in $\textrm{Im}\Sigma_+(\bm{0},\omega)$
for positive and negative energies, corresponding to
the terms (II) and (III) in Eq.~(\ref{eq:imsig}), respectively.
It is seen that ${\rm Re}\Sigma_+( \bm{p},\omega )$
exhibits a steep rise in the two regions corresponding to these peaks.
We also plot the lines $\omega-|\bm{p}|$, i.e.
the l.h.s. of Eq.~(\ref{eq:disp}), by the
dotted lines in Fig.~\ref{fig:self}.
Since the quasi-dispersion $\omega_+(\bm{p})$ is given by
Eq.~(\ref{eq:disp}),
$\omega_+(\bm{p})$ corresponds to the points of intersection of
the dotted lines and ${\rm Re}\Sigma_+( \bm{p},\omega )$.
We see that the crossing points depend strongly on the shape of 
${\rm Re}\Sigma_+( \bm{p},\omega )$.
Near $\omega/m=0.1$, the rapid change of ${\rm Re}\Sigma_+( \bm{p},\omega)$
causes the quasi-dispersion relation to deviate from the free quark dispersion,
as shown in Fig. \ref{fig:qdr1}.

\begin{figure}
\begin{center}
$T/m=1$ \\
\vspace{-0.5cm} 
\begin{tabular}{cc}
\hspace{-.5cm}
\includegraphics[width=220pt]{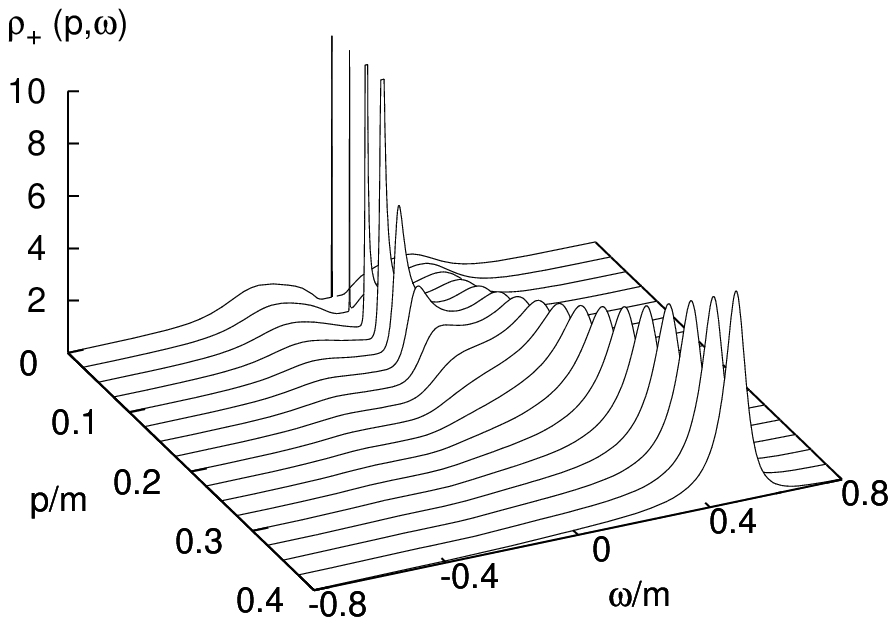} \hspace{-1.8cm} &
\includegraphics[width=240pt]{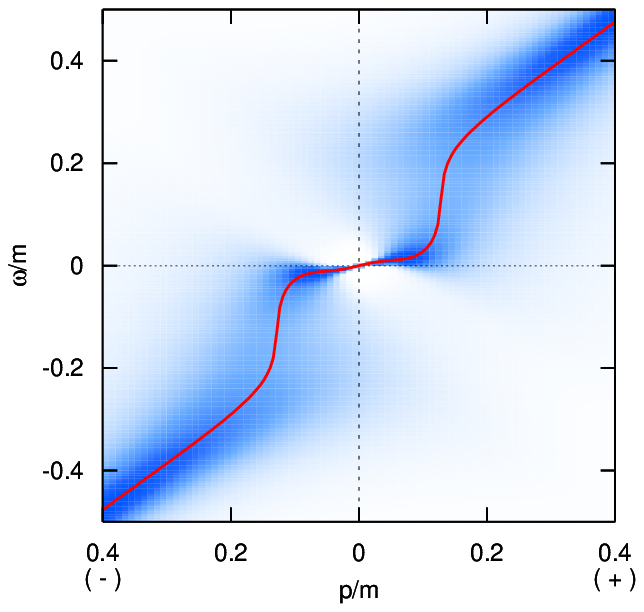}
\end{tabular}
$T/m=1.4$ \\
\vspace{-0.8cm} 
\begin{tabular}{cc}
\hspace{-.5cm}
\includegraphics[width=220pt]{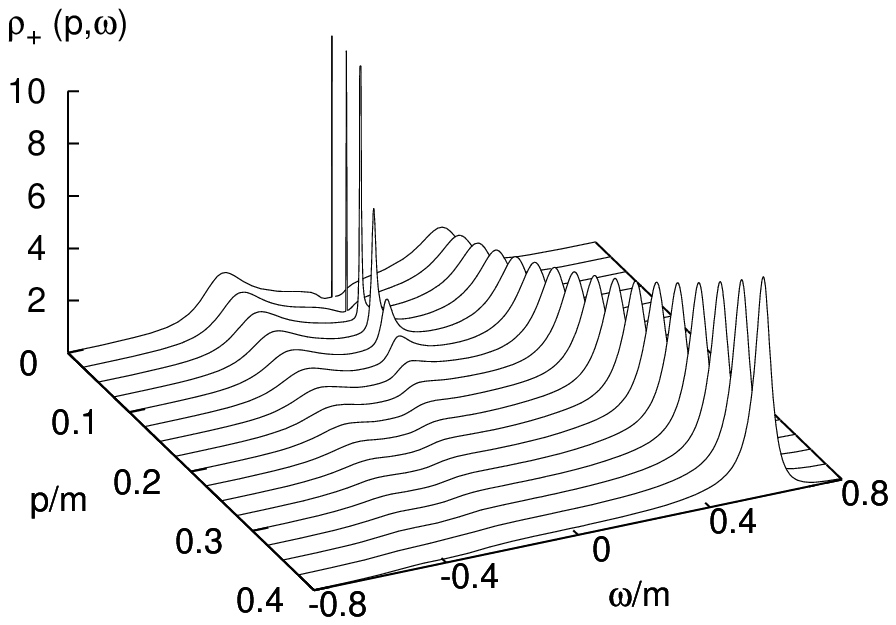} \hspace{-1.8cm} &
\includegraphics[width=240pt]{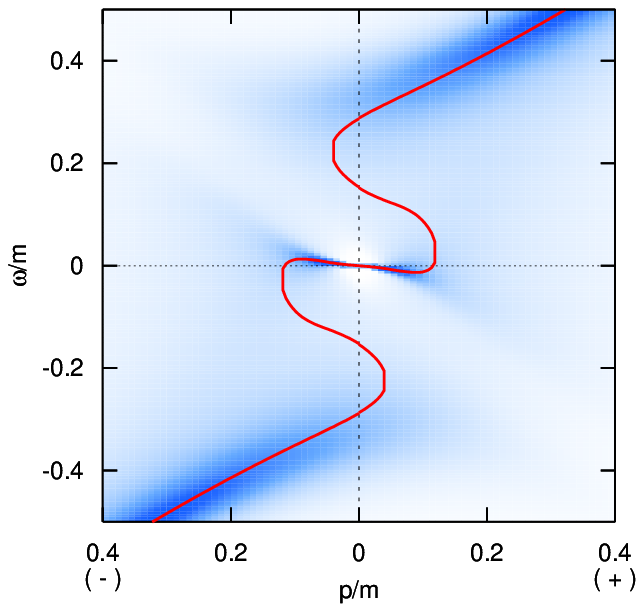}
\end{tabular}
$T/m=1.8$ \\
\vspace{-0.8cm} 
\begin{tabular}{cc}
\hspace{-.5cm}
\includegraphics[width=220pt]{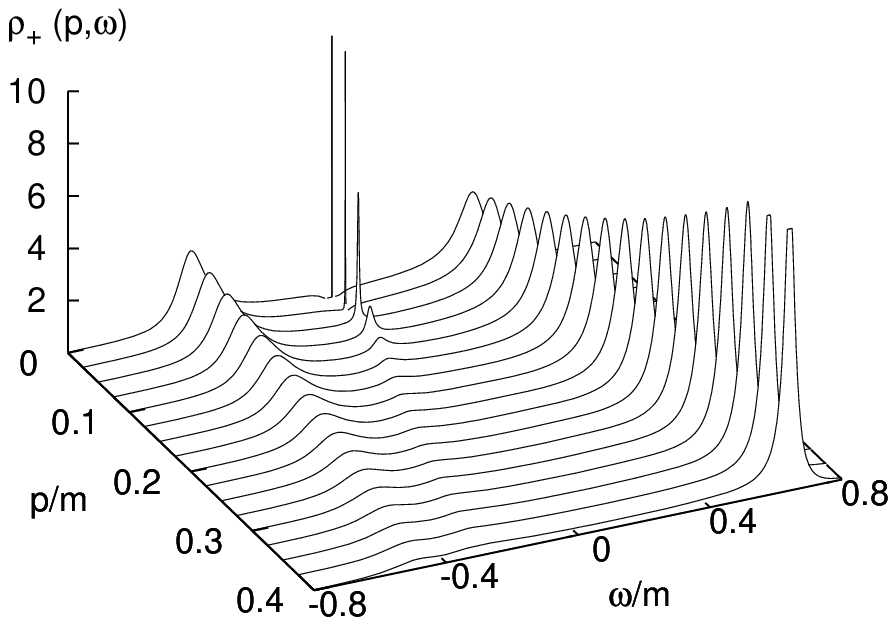} \hspace{-1.8cm} &
\includegraphics[width=240pt]{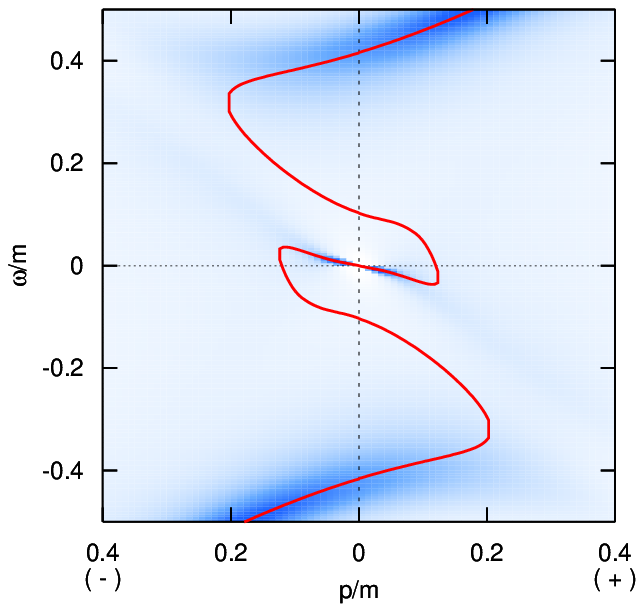}
\end{tabular}
\caption{
The quark spectral function $\rho_+(\bm{p},\omega)$ (left)
and the quasi-dispersion relations $\omega_\pm(\bm{p})$
(right) for $T/m=1.0$, $1.4$,
and $1.8$ from the top.
In the right panels, the $( \omega,\bm{p} )$ dependence
of $\rho_\pm(\bm{p},\omega)$ is represented by the color density.
$\omega_+$ ($\omega_-$) and $\rho_+$ ($\rho_-$) are shown in
the right (left) halves of these panels.
Note the direction of the
momentum scale in the left half-plane is opposite to that of the right
half-plane.
}
\label{fig:spc2}
\end{center}
\end{figure}

\subsubsection{Quark spectral function at intermediate temperature}

As $T$ is raised further so that $T$ becomes comparable with $m$,
$\rho_\pm(\bm{p},\omega)$ and $\omega_\pm(\bm{p})$ 
change drastically; 
note that the bosons become thermally excited 
with a considerable probability at such a temperature.
In the left panels of Fig.~\ref{fig:spc2},
we plot $\rho_+(\bm{p},\omega)$ for $T/m=1.0,\, 1.4$ and $1.8$.
We see that for $T/m=1.0$,
the quasi-particle peak starts to split near $|\bm{p}|/m=0.15$.
Also in this vicinity, there appear broad peaks 
both in the positive and negative energy regions for lower momenta.
Thus the spectral function in the low-momentum region
has \textit{three peaks}.

\begin{figure}[ht]
\begin{center}
\includegraphics[width=0.65\textwidth]{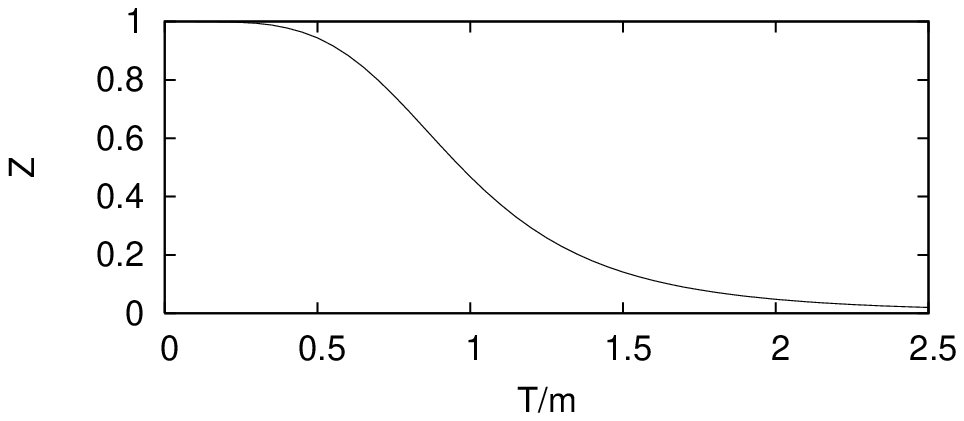} 
\caption{
The temperature dependence of the residue of the pole 
at the origin $(\omega=|\bm{p}|=0$).
}   
\label{fig:residue}
\end{center}
\end{figure}

A remark is in order here. Among the three peaks, 
that near the origin has no width,
and thus it is difficult to read off the strength from 
Fig.~\ref{fig:spc2}.
Here we note that the strength at the origin is actually
given by the residue of the Green functions, $G_\pm$:
The residue is simply given by
\begin{equation}
  Z(T)=[1-\partial\textrm{Re}\Sigma_\pm/\partial\omega]^{-1},
  \label{eq:residue}
\end{equation}
since the imaginary part, ${\rm Im}\Sigma_\pm(\bm{p},\omega)$,
vanish at the origin. 
The $T$ dependence of $Z(T)$ is
shown in Fig.~\ref{fig:residue}.
It is seen that $Z(0)=1$, as mentioned previously, 
and $Z(T)$ decreases gradually as $T$ increases
but is still considerably large for $T/m \sim 1$,
and hence the three-peak structure of the spectral function 
is realized  in this temperature region, as mentioned above.

In the high $T$ limit, $Z(T)$ vanishes,
and only two peaks remain in $\rho_\pm( \bm{p},\omega )$.
As we see below, these two peaks correspond to 
the normal quark quasi-particle and plasmino excitations
obtained in the HTL approximation.

In the right panels of Fig.~\ref{fig:spc2},
we plot the quasi-dispersion relations 
and the $(\omega, p)$ dependence
of the spectral function; the latter is represented by the (color) density.
In these panels, 
$\omega_+$ ($\omega_-$) and $\rho_+$ ($\rho_-$) are shown in
the right (left) halves of the figures. 
Note that the direction of the momentum scale in the left half plane 
is opposite to that in the right half plane.
At $T/m=1$, there is one quasi-dispersion curve for the entire momentum
region.
As $T$ increases, the quasi-dispersion curve bends greatly 
forming `back-bending' parts, and 
there eventually appear multi-valued quasi-dispersion relations
for some momentum regions.
At $T/m=1.4$, the quasi-dispersion relation is
five-valued at low momenta;
note that this back-bending quasi-dispersion relation
is acausal and unphysical.
Only three of these values, however, are accompanied by peaks of 
the spectral function, as
the quasi-dispersion relation
$\omega_+(\bm{p})$ near the back-bending region 
does not form a peak in $\rho_+( \bm{p},\omega )$ and is unphysical.
The back-bending feature of the quasi-dispersion relation becomes
more prominent as $T$ increases further, as shown in
the bottom-right panel in Fig.~\ref{fig:spc2}.

To understand the peculiar behavior of $\rho_+( \bm{p},\omega )$
and $\omega_+(\bm{p})$ displayed in Fig.~\ref{fig:spc2},
we show $\Sigma_+( \bm{p},\omega )$ 
at the intermediate temperature $T/m=1.4$ 
in the right panel of Fig.~\ref{fig:self}.
We see that the two peaks 
in ${\rm Im}\Sigma_+( \bm{p},\omega )$ become sharper
than those for $T/m=0.8$, shown in the left panel, and 
the oscillatory behavior of the real part for
$\omega\sim 0$ also becomes more prominent.
In order to determine the quasi-dispersion relation $\omega_+(\bm{p})$,
i.e., the solutions of Eq.~(\ref{eq:disp}), from the figure,
we plot the line $\omega-|\bm{p}|$, i.e. the l.h.s. of 
Eq.~(\ref{eq:disp}).
In the top panel, we see that there appear five crossing points.
The crossing points with the second and fourth largest $\omega$, however,
are located at the energies of the peak of 
$|{\rm Im}\Sigma_+(\bm{p},\omega)|$, 
and hence the spectral function does not form a peak there.
For large momenta, the number of crossing points decreases, 
and eventually only one crossing point remains.

\subsubsection{Quark spectral function at high temperature}

\begin{figure}[t]
\begin{center}
\begin{tabular}{cc}
\hspace{-.5cm}
\includegraphics[width=220pt]{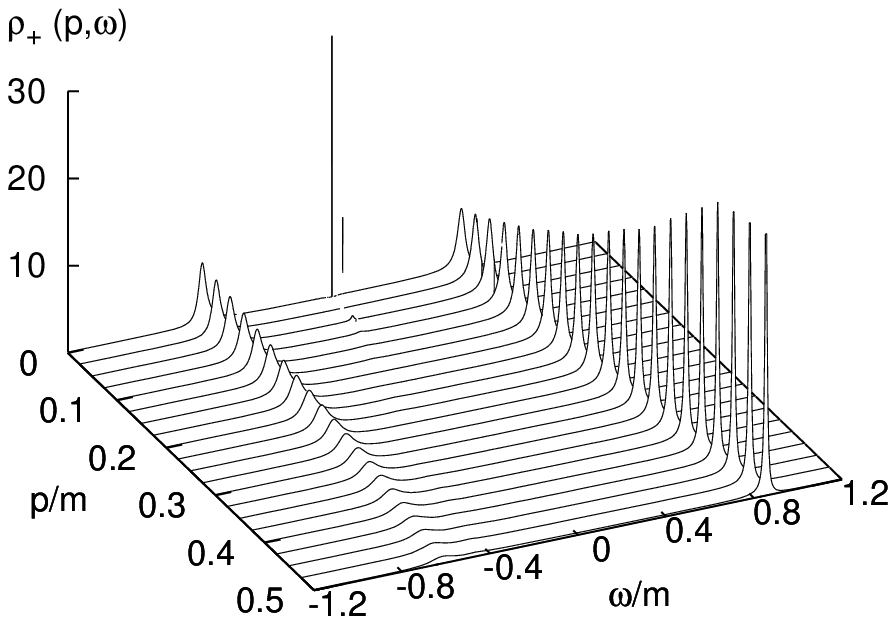} \hspace{-1.5cm} &
\includegraphics[width=240pt]{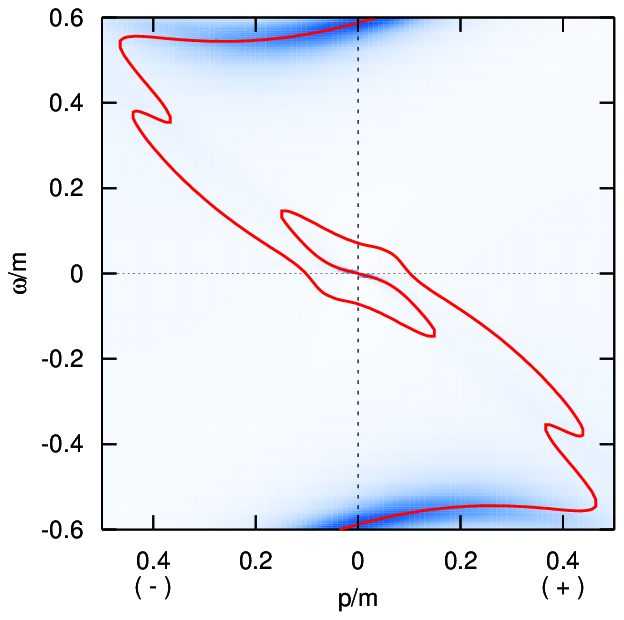}
\end{tabular}
\caption{The quark spectral
function $\rho_+(\bm{p},\omega)$ and
the quasi-dispersion relation $\omega_\pm(\bm{p})$
for $T/m=2.5$.
}
\label{fig:spc3}
\end{center}
\end{figure}
\begin{figure}[h]
\begin{center}
\begin{tabular}{ccc}
\hspace{1cm} $T/m=2$ & \hspace{1cm} $T/m=3$ & \hspace{1cm} $T/m=5$ \\
\hspace{-.5cm}
\includegraphics[width=180pt]{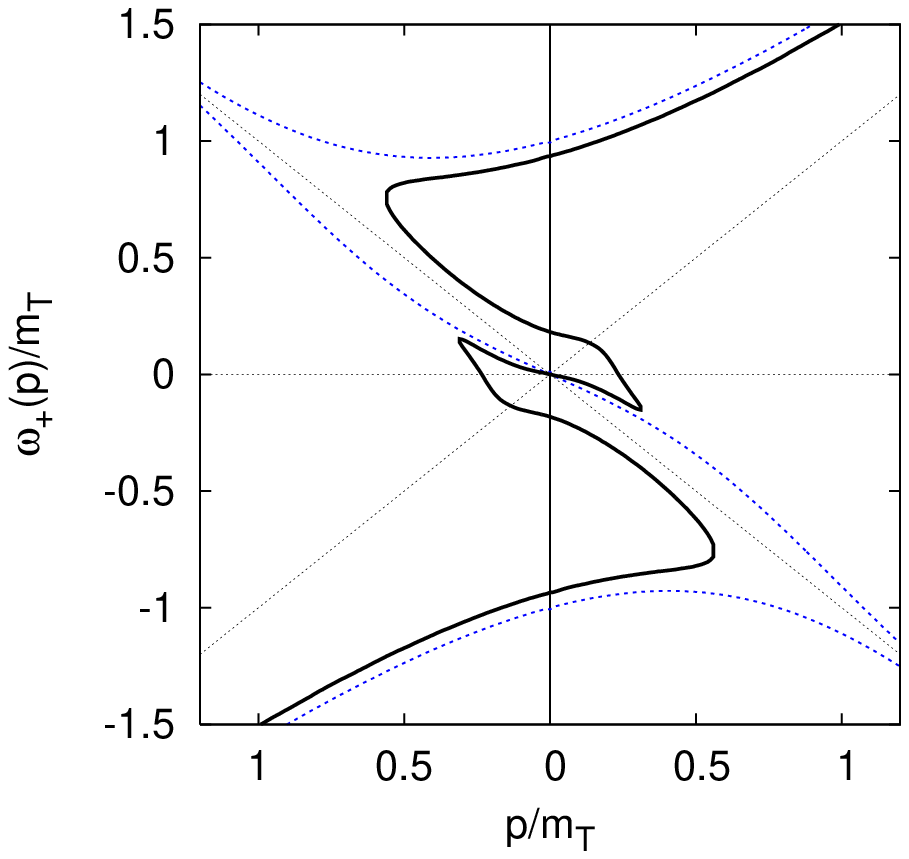} \hspace{-2.1cm} &
\includegraphics[width=180pt]{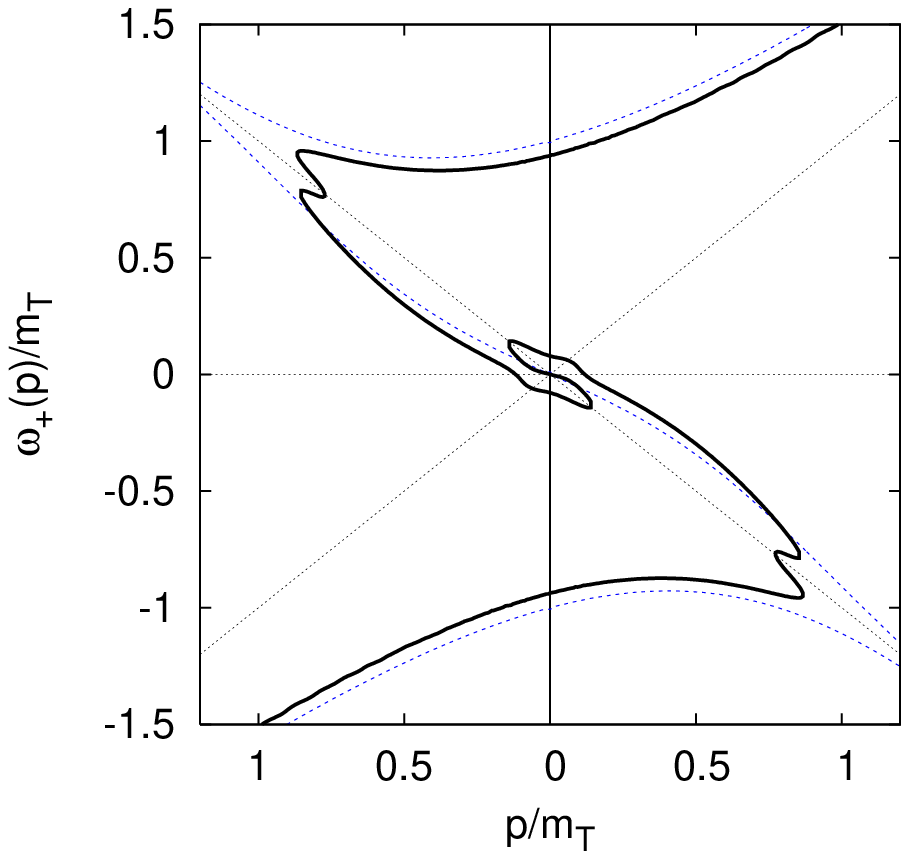} \hspace{-2.1cm} &
\includegraphics[width=180pt]{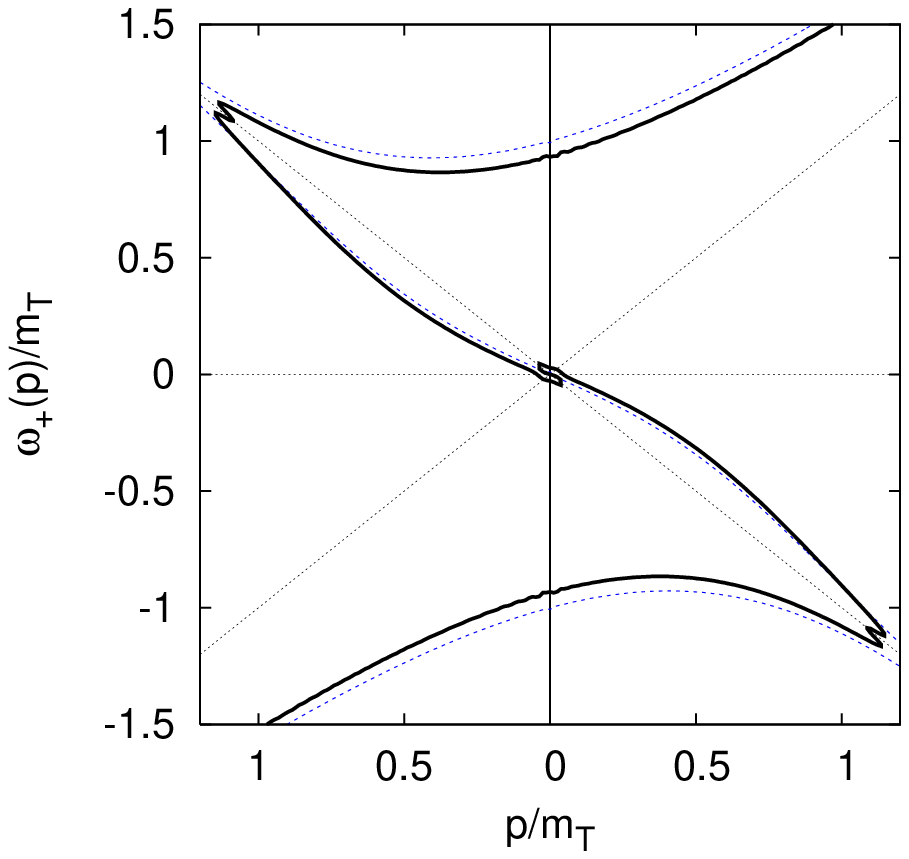} \hspace{-2.1cm}
\end{tabular}
\caption{
The quasi-dispersion relations $\omega_\pm(\bm{p})$ 
for $T/m = 2,3$ and $5$.
All the values are in units of the thermal mass, 
$m_T=gT/4$.
The dashed line represents the dispersion relation in the HTL approximation.
}
\label{fig:HTL}
\end{center}
\end{figure}

Here we examine the quark spectrum at higher temperatures.
We plot $\rho_+(\bm{p},\omega)$ and $\omega_\pm(\bm{p})$
for $T/m=2.5$ in Fig.~\ref{fig:spc3}
as an example of the high-temperature case, i.e. $T\gg m$.
We see that two well-formed peaks appear in the positive and
negative energy regions, while the peak near the origin
immediately disappears as the momentum becomes finite.
The curve representing the quasi-dispersion
relation in the right panel is quite complicated.
It should be noted that, however,
the back-bending behavior of the quasi-dispersion relation is
unphysical, and does not correspond to a peak of the spectral 
function, as mentioned previously.
We plot the quasi-dispersion relation at several temperatures,
$T/m=2,\, 3$ and $5$, in the high-$T$ region
in Fig.~\ref{fig:HTL}, together with that
in the HTL approximation with the thermal mass $m_T=gT/4$.
From the figures,
it is seen that our quasi-dispersion relation approaches that obtained
in the HTL approximation as $T$ increases;
the two curves of $\omega_+(\bm{k})$ approach the quark quasi-particle 
and plasmino
dispersion relations, respectively, with the thermal mass 
$m_T=gT/4$.\footnote{
In fact, the spectrum in Fig.~\ref{fig:HTL}
does not exactly coincide with that in the HTL approximation
because the coupling is not small.
However, it is known that their difference is small at one-loop
order \cite{Peshier:1998dy}.
}

Here we note that there appears 
a continuous spectral bump near $\omega=0$ in the space-like
region in the HTL approximation, 
which is reminiscent of the spectral peak  seen at intermediate
temperatures in the space-like region near $\omega=0$.
We stress, however, that the physical origins of these peaks are
completely different.
The peak appearing at intermediate temperatures is 
due to the pole of the Green function,
as shown in the previous subsection, while 
the bump in the HTL approximation is due to the Landau damping,
and  the quark propagator calculated in this approximation
does not have a pole at $\omega=0$.

\subsection{Discussion of the three-peak structure of the spectral
function for $T\sim m$}

We have seen that the appearance of the three-peak structure  
for values of $T$ comparable with the boson mass $m$ is the most 
characteristic feature
 in the quark spectral function caused 
by the coupling with the massive boson.
In this subsection, we attempt to elucidate
the physical origin of the multi-peak structure.
This discussion proceeds analogously to that\cite{Kitazawa:2005mp}
given for the origin of the three-peak structure caused by the 
soft mode of the chiral phase transition.

To understand the physical mechanism responsible for
the three-peak structure of $\rho_+( \bm{p},\omega )$,
we first recall that 
there develop two peaks in ${\rm Im}\Sigma_+$ 
at a positive and negative energy in this temperature 
region; the peak height here increases with $T$.
As discussed in \S\ref{subsec:selfe},
these two peaks correspond to the decay processes depicted in
(II) and (III) of Fig.~\ref{fig:feynykw}, which are both Landau damping.
The process  (II) is 
the annihilation process of the incoming quark $Q$ and the thermally excited
antiquark into a boson in the thermal bath,\,
$Q+\bar{q}\to b$,\, and  its inverse process.
Two remarks are in order here.
First, the disappearance of an anti-quark 
implies the creation of a `hole' in the 
thermally excited anti-quark distribution\cite{Weldon:1989ys}.
Second, the creation of bosons in a thermal bath
is enhanced in comparison with the
case in vacuum by a  statistical factor of $1+n$, 
which becomes large when $T$ is comparable to $m$. 
Thus, we see that the process (II) causes a virtual mixing 
between the quark and `anti-quark hole' states through
the coupling with the boson in a thermal bath,
\footnote{
We remark that quark number conservation is not violated with this
mixing, because an `anti-quark hole' has a positive quark number.}
and as a result, the mixing is enhanced when $T/m\sim 1$.

The process (III) is another decay process of a quasi-quark
state $Q$, which is now understood to be
a mixed state of quarks and antiquark-holes,
into an on-shell quark 
via a collision with a thermally excited boson:
$Q+ b \to q$ and its inverse process,
$q \to Q+ b$.
As is seen from the left panel of Fig.~\ref{fig:imsig},
the energy involved in probable processes occurring
at small momenta is negative for the  process (III).
These processes again give rise to a mixing of a quasi-quark and
an anti-quark hole state.
As seen from the Bose-Einstein distribution function $n$,
thermally excited bosons are abundant when 
$T$ approaches $m$.

A similar interpretation applies to the anti-quark sector,
${\rm Im}\Sigma_-(\bm{p},\omega)$,
if the quark and anti-quark hole are replaced by 
an anti-quark and a quark hole,  respectively.
Thus, the process corresponding to (II) in Fig.~\ref{fig:feynykw} 
is $\bar{Q}+q\to b$, where $q$ denotes a thermally
excited quark, for instance.
For this reason, from this point we only consider the quark sector.

\begin{figure}[t]
\begin{center}
\includegraphics[width=140pt]{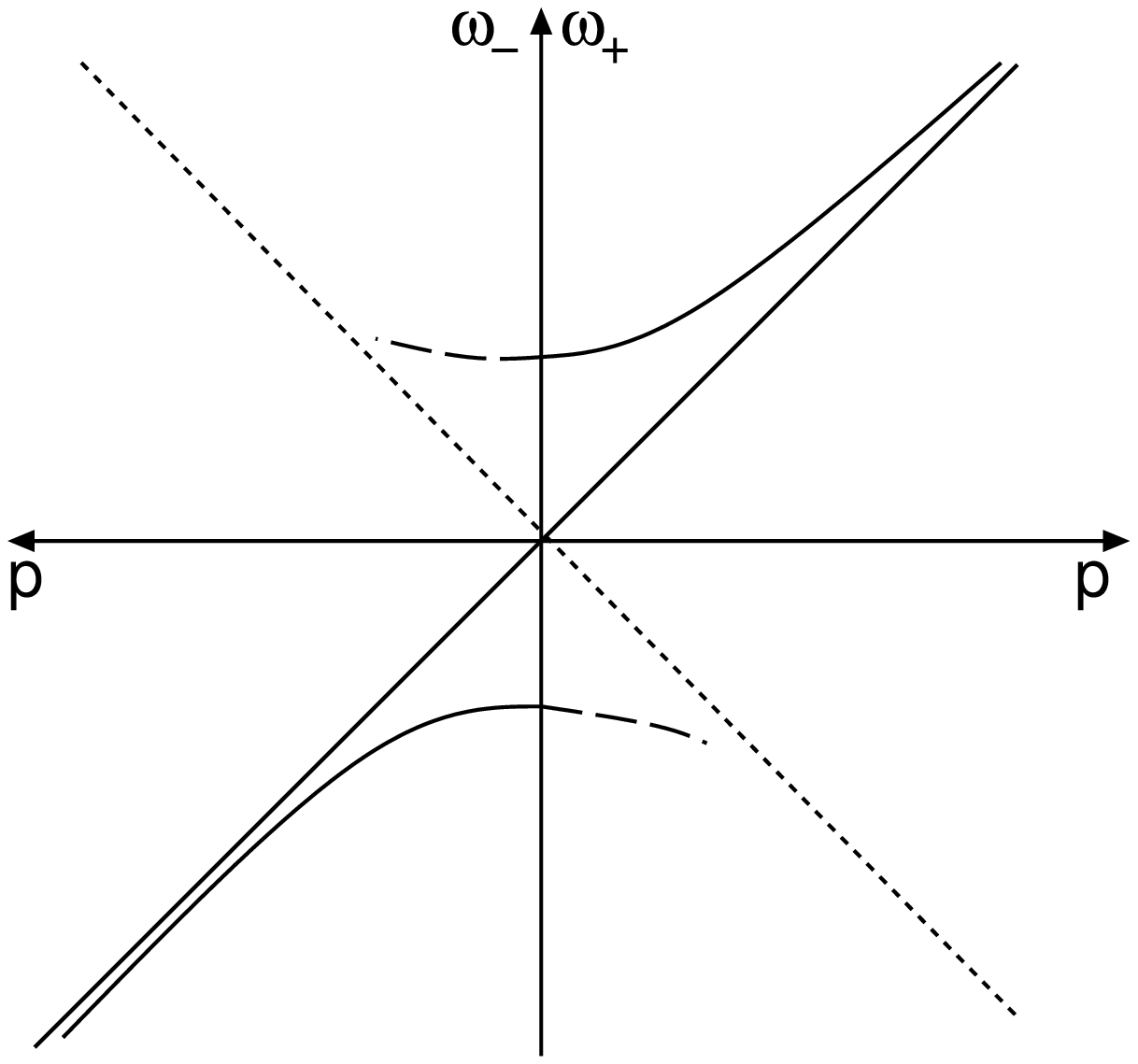}
\hspace{5mm}
\includegraphics[width=140pt]{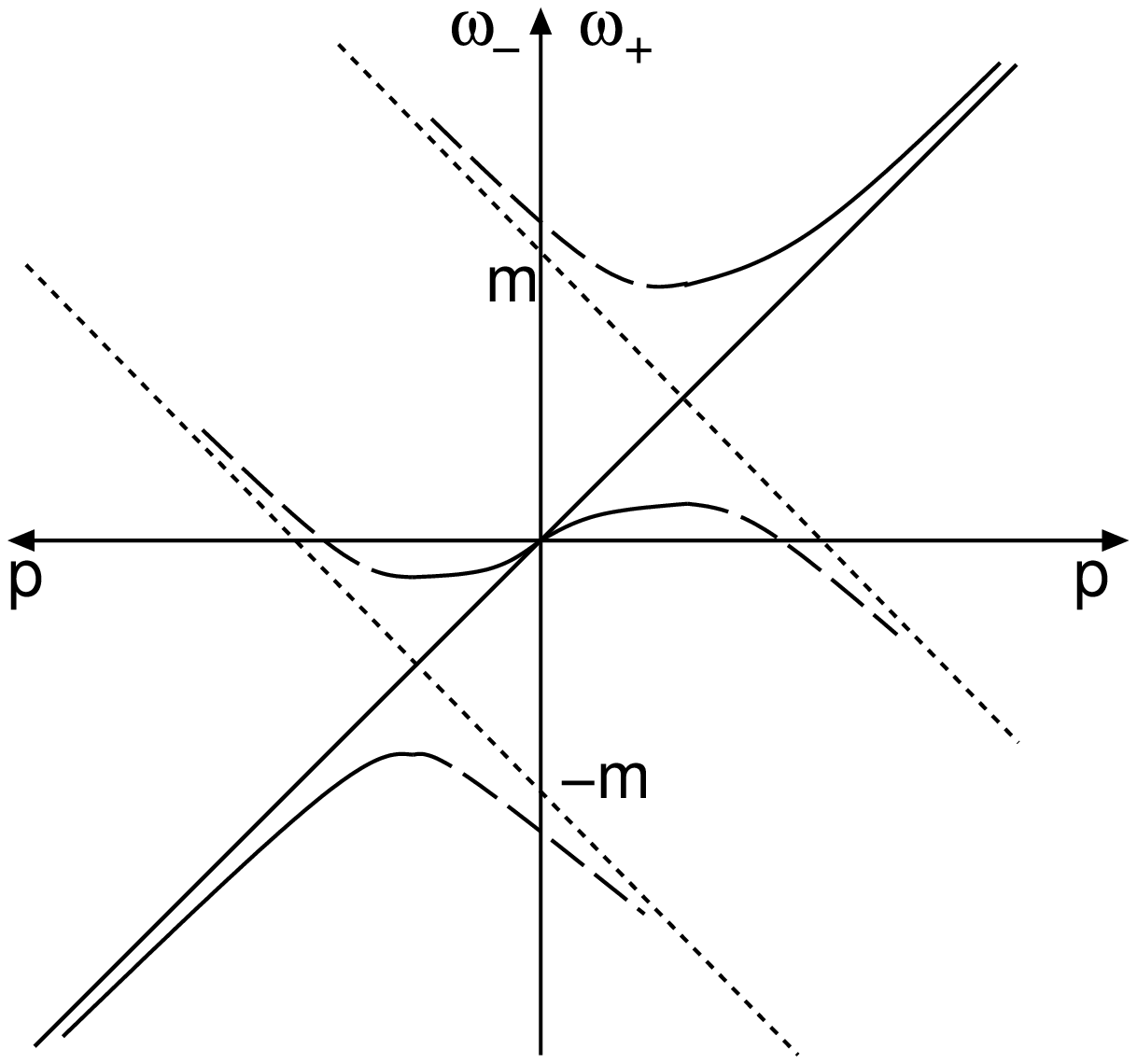}
\caption{Typical peak position of the
spectral functions in the case of level mixing at the
origin (left) and away from the origin (right).
 The long dashed curves show that the strength of the spectrum
is becoming weaker.
The solid line represents the free quark and antiquark 
dispersion relations, and
the dotted line the antiquark hole and quark hole dispersion
relations.
Level repulsion takes place at the intersection point of these two lines.}
\label{fig:reso}
\end{center}
\end{figure}

The mechanism for the mixing of the quark and hole state 
can also be characterized as a {\em resonant scattering}
\cite{Janko1997},
which was originally introduced to understand the non-Fermi liquid 
behavior of fermions just
above the critical temperature of the superconducting
transition\cite{Janko1997,Kitazawa:2005pp}.
In fact, we have seen 
that the process (II) in Fig.~\ref{fig:feynykw}
includes a scattering process of
the  quark  by a massive boson, thus creating
a hole state in the thermally distributed anti-quark states:
$Q\to \bar{q}_h\, +\, b$.
Such a process is called resonant scattering\cite{Janko1997}.
Note that the most probable
process for finite $T$ involves the lowest energy state of the boson,
 i.e., a rest boson with a energy $m$. The energy conservation law
in the most probable case for the above process is
$\omega_{Q}(\bm{p})+\omega_{\bar{q}}(-\bm{p})= m$, or equivalently,
\beq
\omega_{Q}(\bm{p})= m\, -\, \omega_{\bar{q}}(-\bm{p}).
\eeq
This equation actually represents the energy-momentum relation for the
particles involved in the state mixing. Thus we see that the
physical energy spectrum is obtained as a result of the level
repulsion between the energies 
$\omega_{q}(\bm{p})=\vert \bm{p}\vert$ and 
$m\, -\, \omega_{\bar{q}}(-\bm{p})=m-\vert \bm{p}\vert$ 
in the perturbative picture.
This situation is depicted in the upper-right part 
of the right panel in Fig.~\ref{fig:reso}.

Similarly, the process (III) in Fig.~\ref{fig:feynykw}
includes the process
$q\to Q +b$, and the energy-momentum conservation law for
this process for the most probable case is
\beq
-\omega_{q}(\bm{p})= -m\, + \omega_{Q}(\bm{p}).
\eeq
Thus, the physical energy spectrum is obtained as a result of the level
repulsion between the energies 
$-\vert \bm{p}\vert$ and $-m +\vert \bm{p}\vert$.
This situation is also 
depicted in the lower-right part of the right panel
in Fig.~\ref{fig:reso}.

We thus find that at temperatures satisfying $T/m \sim 1$,
owing to the finite boson mass,
 the level repulsions occur far from the origin, and then 
the quasi-dispersion relations are bent twice,
or two gap-like structures in the quark spectrum are formed
at positive and negative energies,
 as shown in the right panel of Fig.~\ref{fig:reso}.

It is interesting to  consider 
the high temperature limit, $T\gg m$, or $m/T \sim 0$.
In this case, the effect of the boson mass can be ignored,
and the resonant scattering occurs only once at the origin
($\omega=|\bm{p}|=0$),
since the energy levels which are to be repelled
cross only there.
Then 
the situation becomes that represented in the left panel of 
Fig. \ref{fig:reso}.

It has been shown \cite{Kitazawa:2005mp} that 
the quark spectral function possesses a 
three-peak structure near $T_c$ of the chiral transition
 when the chiral soft mode\cite{Hatsuda:1985eb}
is incorporated into the quark self-energy.
As was described in the Introduction,
the soft modes behave like a massive elementary 
boson with a mass $m_\sigma^*(T)$ as $T$ approaches $T_c$, i.e.
$\omega_{\rm soft}\sim \sqrt{\bm{p}^2+m^*_{\sigma}(T)^2}$,
and hence the quark spectra studied in Ref.~\citen{Kitazawa:2005mp}
are essentially the same as that treated in this section.
We also note that 
as $T$ is lowered toward $T_c$, $m_\sigma^*(T)$ tends to vanish,
and hence the ratio $T/m_\sigma^*(T)$ becomes large.
Thus the quark spectrum
approaches that in the $T/m\to\infty$ limit of the present work
\cite{Kitazawa:2005mp}.

\subsection{Quark spectral function for various coupling constants}
\label{subsec:g}

\begin{figure}[t]
\begin{center}
\begin{tabular}{cc}
\hspace{-1.8cm} $g=0.3$ , $T/m=2.8$ & \hspace{-2.3cm} $g=2$ , $T/m=1$ \\
\hspace{-.7cm}
\includegraphics[width=220pt]{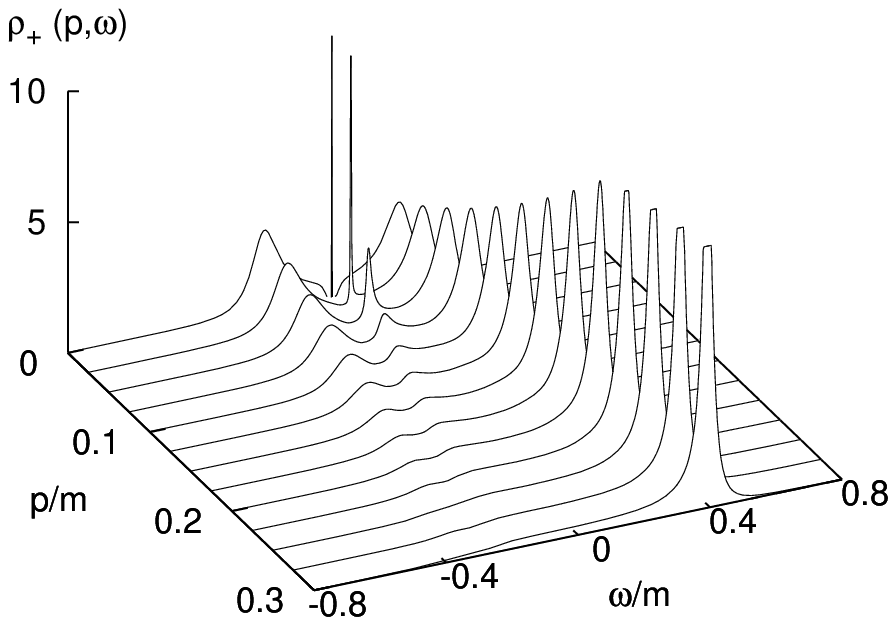} & \hspace{-1.3cm}
\includegraphics[width=220pt]{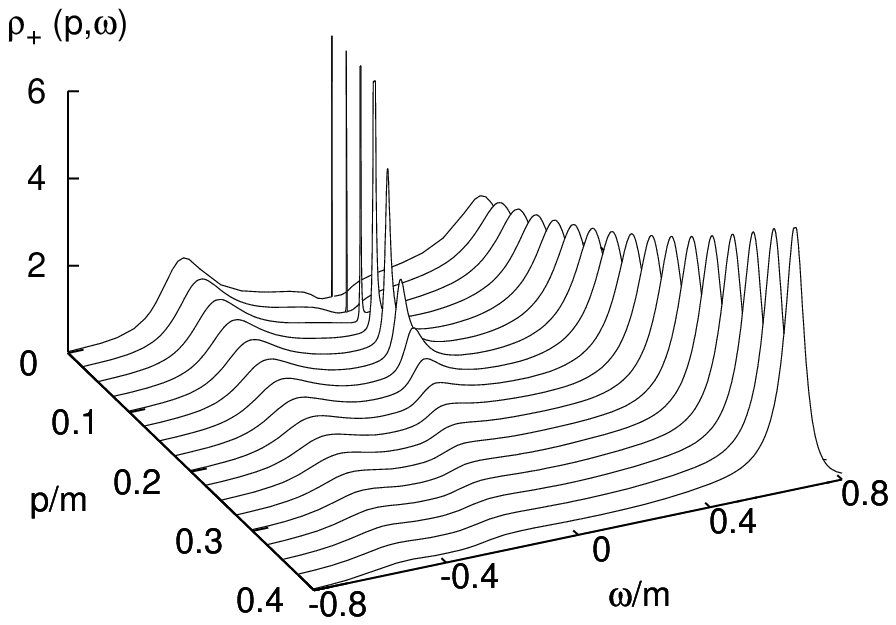} 
\end{tabular}
\caption{
The quark spectral functions $\rho_+(\bm{p},\omega)$
for $g=0.3$, $T/m=2.8$ and $g=2$, $T/m=1$.
}   
\label{fig:spc_g}
\end{center}
\end{figure}

We have derived the quark spectrum with the fixed coupling $g=1$,
and we have found novel three-peak structure 
at intermediate temperatures.
How do these features change as $g$ is varied?
In the previous subsections, we pointed out that
the Landau damping, which induces the two peaks
in ${\rm Im}\Sigma^R ( \bm{p},\omega )$,
plays an essential role in creating
the three-peak structure in the quark spectrum.
Because $\Sigma^R ( \bm{p},\omega )/g^2$
is independent of $g$ at one-loop order,
we conjecture that the three-peak structure in
the quark spectrum would not be affected qualitatively 
even if we varied the coupling $g$.
We confirmed numerically that this is indeed the case:
There appears similar three-peak structure in the quark spectrum
for all values of $g$ that we considered, although the temperature at which 
the clear three-peak structure appears depends on $g$.

In Fig.~\ref{fig:spc_g}, we plot the quark spectral function 
$\rho_+ (\bm{p},\omega)$ with $g=0.3$ and $2$, for values of $T$ where
a clear three-peak structure is seen.
The spectral function possesses a single-peak structure at values of $T$
lower than in the case considered in Fig.~\ref{fig:spc_g}, 
and it approaches the two-peak structure at higher $T$.
It is notable that for larger $g$,
the three-peak structure appears
at lower $T$ and the distance between the peaks becomes larger.

\section{Quark spectrum in a Yukawa model with a massive vector
(axial-vector)  boson}
\label{sec:vector}

In this section, we investigate the quark spectrum using a Yukawa model
with a massive vector (axial-vector)
 boson  and show that the  three-peak structure
of the quark spectral function is obtained.

We start from the following Lagrangian, composed of 
a massless quark $\psi$ and vector boson field $V_\mu$:
\begin{eqnarray}
{\cal L}
= \bar\psi ( i\partial\sla - ig \gamma^\mu V_\mu ) \psi + {\cal L}_{\rm v}.
\label{eq:LagrangianVector}
\end{eqnarray}
We consider the following simplest form for the Lagrangian of the 
massive vector field, ${\cal L}_{\rm v}$:
\begin{eqnarray}
{\cal L}_{\rm v}
= - \frac14 F_{\mu\nu}F^{\mu\nu} + \frac12 m^2 V_\mu V^\mu,
\label{eq:LagrangianV}
\end{eqnarray}
with the field strength 
$ F_{\mu\nu} = \partial_\mu V_\nu - \partial_\nu V_\mu $.

As in the case for the scalar boson, we find that the replacement
of the vector boson with an axial-vector boson does not lead
to any difference in the quark spectrum in the perturbation theory
in this work.

\subsection{Quark self-energy}

At one-loop order, 
the quark self-energy in the imaginary time formalism is 
given by
\begin{eqnarray}
\tilde\Sigma ( \bm{p} , i\omega_m )
= -g^2 T \sum_n \int \frac{ d^3 \bm{k} }{ (2\pi)^3 }
\gamma^\mu {\cal G}_0( \bm{k},i\omega_n ) \gamma^\nu 
{\cal D}_{\mu\nu} ( \bm{p}-\bm{k} , i\omega_m-i\omega_n ),
\label{eq:SigmaVImag}
\end{eqnarray}
with the Matsubara propagator for the massive vector boson,
\begin{eqnarray}
{\cal D}_{\mu\nu}( i\nu_n , \bm{p} )
= - \frac{ g_{\mu\nu} - \tilde p_\mu \tilde p_\nu / m^2 }
{ \tilde p_\mu \tilde p^\mu - m^2 },
\label{eq:Proca}
\end{eqnarray}
where $ \tilde p_\mu = ( i\nu_n , \bm{p} ) $.
The propagator for this vector boson,
which is obtained from Eq.~(\ref{eq:LagrangianV}),
is called the Proca propagator\cite{Proca}. 
It is known that the massless limit
of Eq.~(\ref{eq:Proca}) cannot be taken, 
because of the lack of gauge invariance of Eq.~(\ref{eq:LagrangianV}).
As we see below, this property causes a problem
in the the high temperature limit, $T/m\to\infty$,
because this limit also represents the massless limit
with fixed $T$.
One should describe a boson field in the Stuckelberg formalism,\cite{Proca}
which has a massless limit, if we study the spectrum at high $T$.
In this work, because we explore the qualitative effect
of the phenomenological massive vector boson on the quark spectrum 
at intermediate $T$, we do not consider this problem
associated with the Proca propagator.

 For coupling with an axial-vector boson, the self-energy
has the same form as Eq.~(\ref{eq:SigmaVImag}), because the $\gamma_5$ matrices
in the vertices cancel out in the case of a massless quark propagator 
${\cal G}_0$.
Therefore, the following results hold also for the coupling with an 
axial-vector boson, as mentioned above.

Carrying out the summation over the Matsubara modes
in Eq.~(\ref{eq:SigmaVImag}) and applying the analytic continuation
$ i\omega_n \to \omega + i\eta $
(see Appendix~\ref{app:vector} for details),
we obtain
\begin{eqnarray}
\Sigma^R( \bm{p},\omega ) 
&=&
g^2 \sum_{s,t=\pm} t \int \frac{ d^3 \bm{k} }{ (2\pi)^3 }
\frac { \gamma^\mu \Lambda_s (\bm{k}) \gamma^0 \gamma^\nu }{2E_b} 
\left( g_{\mu\nu} - \frac{ q_\mu q_\nu }{m^2} \right)
\frac{ f( sE_f ) + n( -tE_b ) }{ \omega + i\eta - sE_f - tE_b }, \notag\\
\label{eq:SigmaVector}
\end{eqnarray}
with $q_\mu = ( \omega - E_f , \bm{p} - \bm{k} )$.
In order to eliminate the ultraviolet divergence in Eq.~(\ref{eq:SigmaVector}),
we employ the same strategy as in the previous section:
We first divide Eq.~(\ref{eq:SigmaVector}) 
into $T$-independent and $T$-dependent parts as
$\Sigma^R ( \bm{p},\omega ) 
= \Sigma^R ( \bm{p},\omega )_{T=0}+ \Sigma^R ( \bm{p},\omega )_{T\ne0}$
and calculate $\Sigma^R( \bm{p},\omega )_{T=0}$ 
while imposing the renormalization condition.

In the present case, however, unlike in the previous section,
the $T$-independent part, $\Sigma^R ( \bm{p},\omega )_{T=0}$, 
includes a divergence in the term
proportional to $p\sla^3$, and the on-shell renormalization condition alone
cannot eliminate the divergence (see Appendix~\ref{app:T=0}).
This is due to the bad ultraviolet behavior coming from the
Proca propagator $D^R_{\mu\nu}$.
Here we impose the renormalization condition 
$ \partial^3 \Sigma^R( p ) / \partial p\sla^3 |_{p\sla=0} = 0 $
to eliminate this divergence and obtain the following renormalized form:
\begin{eqnarray}
\Sigma^R(p)_{T=0}
= \frac{g^2}{32\pi^2} p\sla
\left( \frac{ P^2 + 2m^2 }{m^2} \frac{ (P^2-m^2)^2 }{P^4}
\log \left| \frac{ m^2-P^2 }{ m^2 } \right|
-\frac56 \frac{P^2}{m^2} -2 + 2 \frac{m^2}{P^2} \right)
\nonumber \\
-i\frac{g^2}{32\pi^2} p\sla 
\frac{ P^2 + 2m^2 }{m^2} \frac{ (P^2-m^2)^2 }{P^4}
\epsilon(p_0) \theta( P^2-m^2 ).
\nonumber \\
\label{eq:SigmaVectorT=0}
\end{eqnarray}
The $T$-dependent part $\Sigma^R( \bm{p},\omega )_{T\ne0}$ 
can be calculated
using the same procedure as in the previous section;
i.e., we first calculate the imaginary part 
${\rm Im}\Sigma^R ( \bm{p},\omega )_{T\ne0}$
and then derive the real part 
using the dispersion relation, Eq.~(\ref{eq:Kramers-Kronig}).

From Eq.~(\ref{eq:SigmaVector}),
we obtain
\begin{eqnarray}
  {\rm Im}\Sigma^R ( \bm{p},\omega ) &=&
    -\pi g^2 \int \frac{ d^3 \bm{k} }{ (2\pi)^3 }
    \frac { \gamma^\mu \Lambda_s (\bm{k}) \gamma^0 \gamma^\nu }{2E_b} 
    \left( g_{\mu\nu} - \frac{ q_\mu q_\nu }{m^2} \right) \nonumber \\
  && \times
    \{ -(1+n-f) \delta ( \omega - E_f - E_b )
    -(n+f) \delta(\omega+E_f-E_b) \nonumber \\
  && + (n+f) \delta(\omega-E_f+E_b) 
     + (1+n-f) \delta(\omega+E_f+E_b) \},
  \label{eq:ImSigmaVector}
\end{eqnarray}
with $n=n(E_b)$ and $f=f(E_f)$.
The four terms in Eq.~(\ref{eq:ImSigmaVector}) 
have the same statistical factors and delta functions
as those in Eq.~(\ref{eq:imsig}),
which means that 
the decay processes of the quasi-particles described by these terms
can be understood diagrammatically from Fig.~\ref{fig:feynykw}.
The region in the energy-momentum plane where each term has a finite value
is the same as in the case of Fig.~\ref{fig:imsig}.

For $\bm{p}=\bm{0}$, we obtain
\begin{eqnarray}
{\rm Im}\Sigma_\pm ( \bm{p}=0,\omega )
&=&
- \frac{ g^2 }{ 32\pi }
\frac{ \omega^2 + 2m^2 }{ m^2 }
\frac{ ( \omega^2-m^2 )^2 }{ \omega^3 }
\left( \coth \frac{ \omega^2+m^2 }{ 4T\omega }
+ \tanh \frac{ \omega^2-m^2 }{ 4T\omega } \right). \notag \\
\label{eq:ImSigP_p0}
\end{eqnarray}
Equation~(\ref{eq:ImSigP_p0}) differs from Eq.~(\ref{eq:ImSigY_p0_simple})
in the Yukawa model with a scalar boson
by a factor of $ ( \omega^2 + 2m^2 )/m^2 $.
For small energy, i.e. $\omega \lesssim m$, 
this term approximately gives an overall factor of $2$.
Therefore, we expect that the two-peak structure 
of ${\rm Im}\Sigma_\pm ( \bm{p}=0,\omega )$ caused by the Landau damping
at $|\omega|<m$ and, hence,
the three-peak structure in the quark spectral function
is not qualitatively altered
even in the present case.
For large $\omega$, however, the factor behaves 
as $\omega^2/m^2$, 
which makes Eq.~(\ref{eq:ImSigP_p0}) much  larger
than Eq.~(\ref{eq:ImSigY_p0_simple}) for $|\omega|\gg m $.
This leads to a difference between the scalar and the vector bosons at
very high $T$.

\subsection{Quark spectrum at intermediate temperatures}

\begin{figure}[t]
\begin{center}
\includegraphics[width=0.66\textwidth]{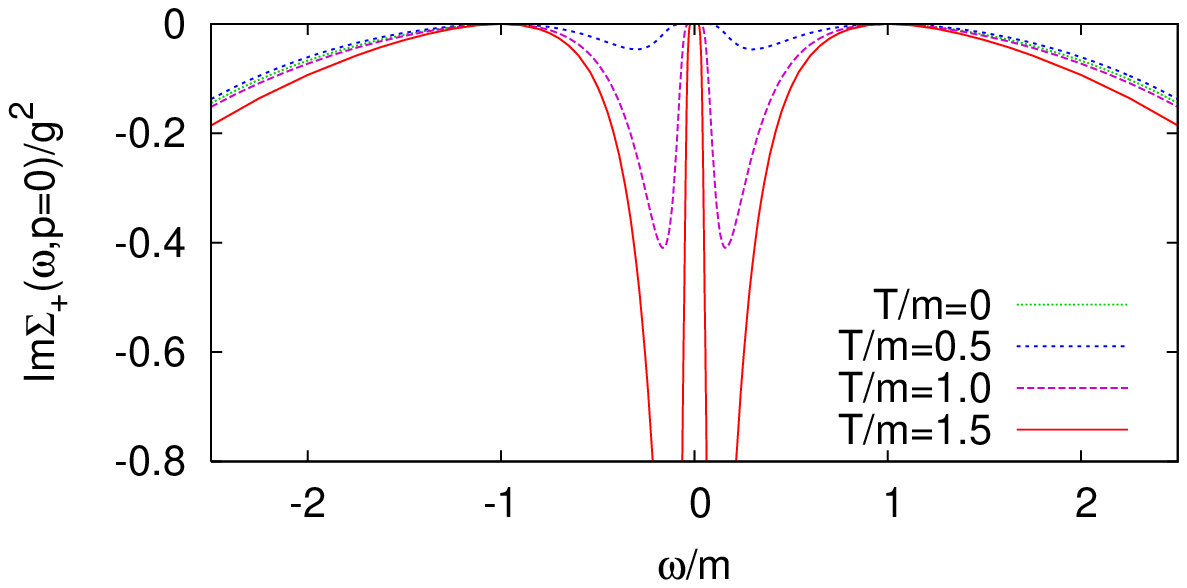}
\includegraphics[width=0.66\textwidth]{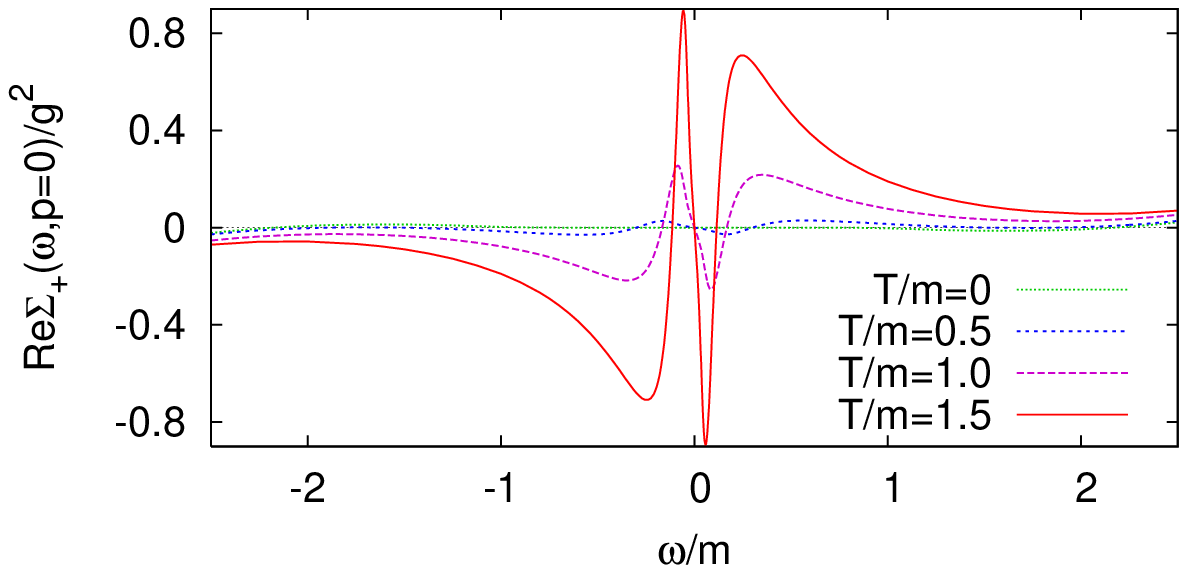}
\caption{
The imaginary and real parts of the self-energy, $\Sigma_+ (\bm{p},\omega)$,
at zero momentum
for several temperatures.
}
\label{fig:ImSigmaP_p0}
\end{center}
\end{figure}

In Fig.~\ref{fig:ImSigmaP_p0},
we plot the imaginary and real parts of $\Sigma_+ (\bm{p},\omega )$ 
for $\bm{p}=0$ at $T/m=0,0.5,1$ and $1.5$.
We fix the coupling constant at $g=1$.
As can be deduced from the above discussion,
the qualitative features of both parts for $|\omega|/m<1$ are
quite similar to those in Fig.~\ref{fig:ImSigma_p0}:
There are two clear peaks in ${\rm Im}\Sigma_+ $ for $|\omega|/m<1$ 
and oscillating behavior in ${\rm Re}\Sigma_+ (\bm{p}=0,\omega)$,
and they grow rapidly as $T$ increases.
For $|\omega|/m>1$, it is seen that 
$|{\rm Im}\Sigma_+(\bm{p}=0,\omega)|$ grows more rapidly 
than in the Yukawa model with the scalar boson shown in 
Fig.~\ref{fig:ImSigma_p0}.

\begin{figure}[t]
\vspace{-5mm}
\begin{center}
\includegraphics[width=0.65\textwidth]{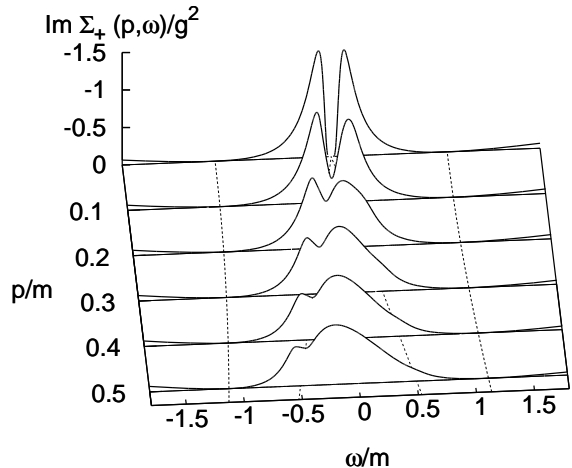}
\vspace{-5mm}
\caption{
The imaginary part of the self-energy, $|{\rm Im}\Sigma_+ (\bm{p},\omega)|$, 
for $T/m=1.5$.
}
\label{fig:ImSigma3dP}
\end{center}
\end{figure}

In Fig.~\ref{fig:ImSigma3dP},
we plot $|{\rm Im}\Sigma_+ (\bm{p},\omega)|$ 
in the energy-momentum plane for $T/m=1.5$.
We see that the qualitative features of 
$|{\rm Im}\Sigma_+ (\bm{p},\omega)|$ are again almost the same 
as those in Fig.~\ref{fig:ImSigma3d} for $|\omega|/m \lesssim 1$.

\begin{figure}[t]
\vspace{-5mm}
\begin{center}
\begin{tabular}{cc}
\hspace{-1.8cm} $T/m=0.4$ & \hspace{-2.3cm} $T/m=0.8$ \\
\hspace{-.7cm}
\includegraphics[width=213pt]{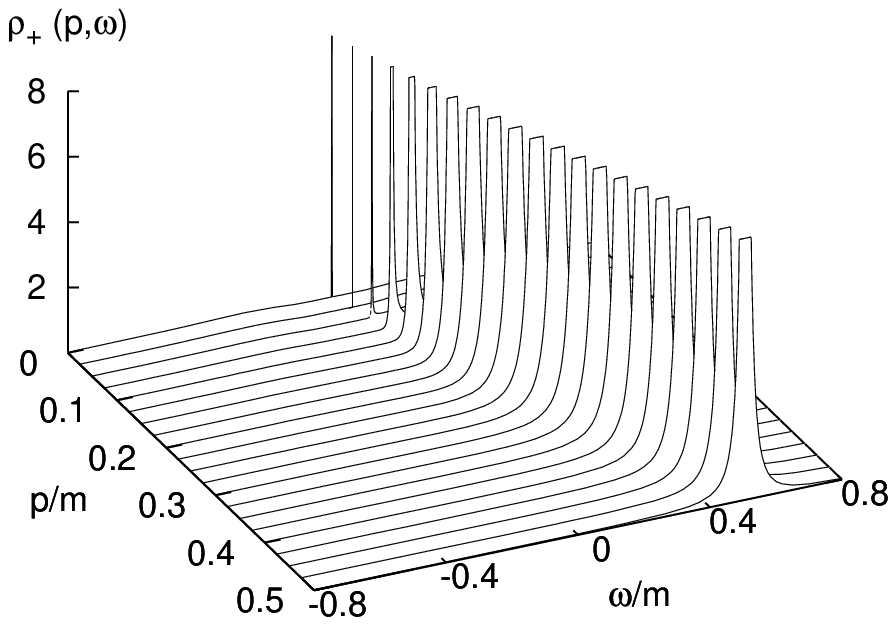} & \hspace{-1.3cm}
\includegraphics[width=213pt]{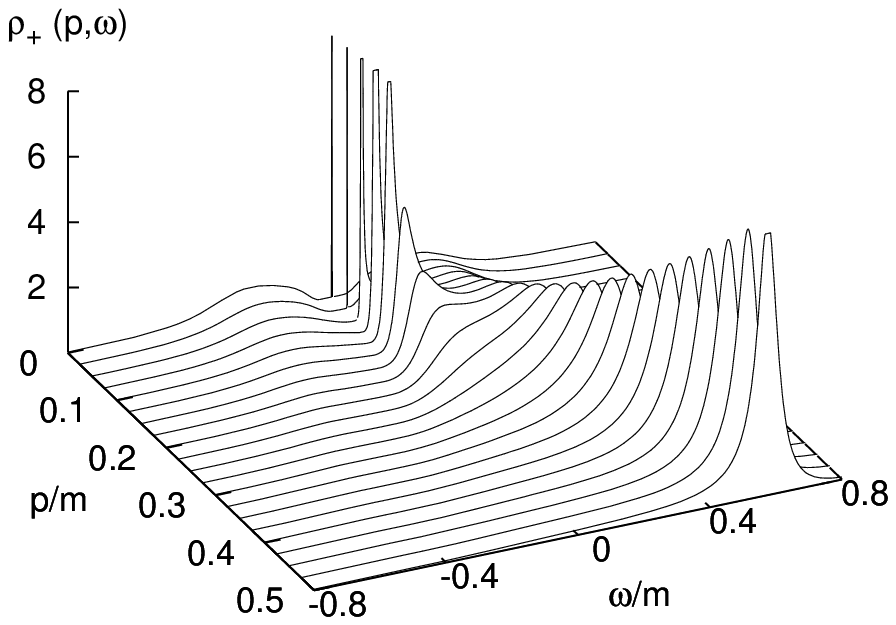} \\
\hspace{-1.8cm} $T/m=1$ & \hspace{-2.3cm} $T/m=1.2$ \\
\hspace{-.7cm}
\includegraphics[width=213pt]{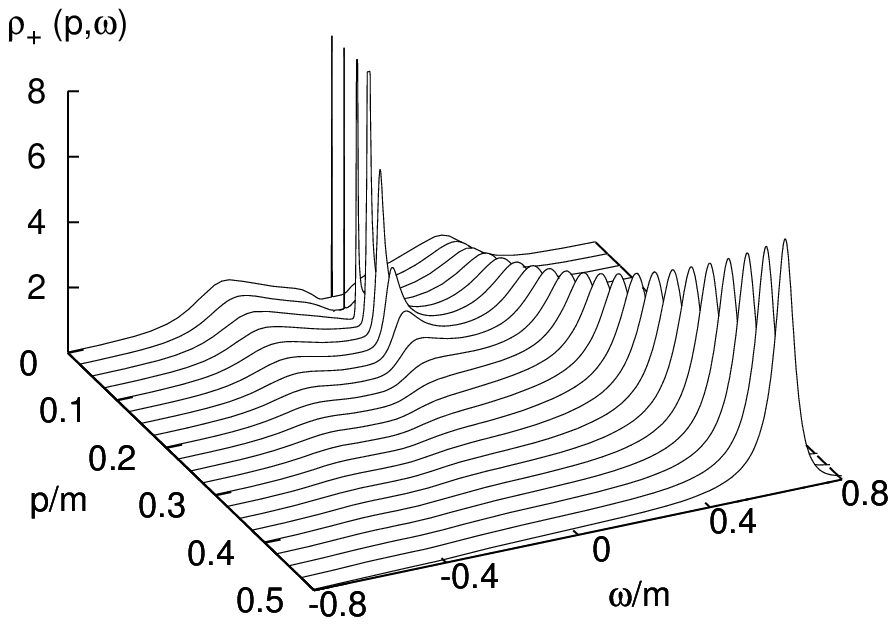} & \hspace{-1.3cm}
\includegraphics[width=213pt]{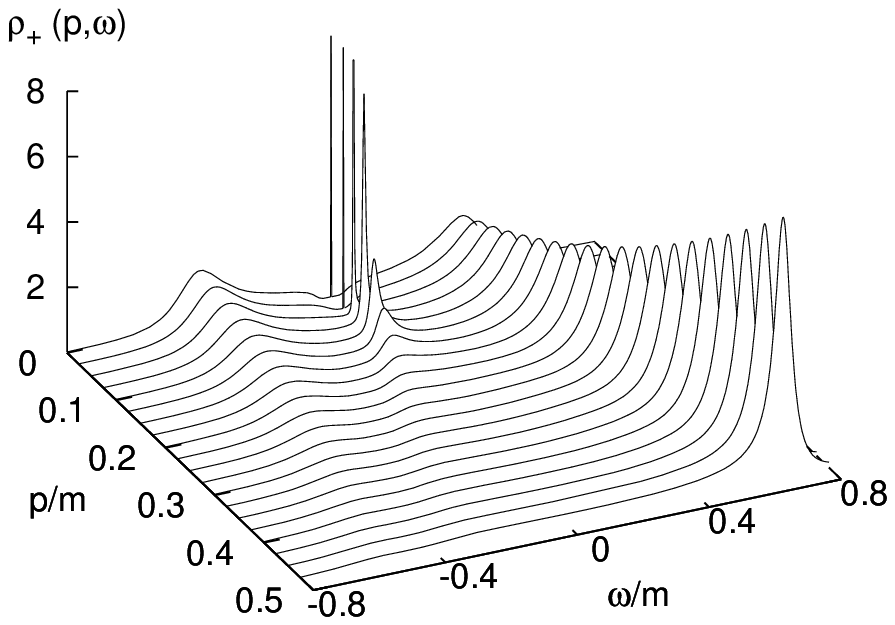} \\
\end{tabular}
\caption{The quark spectral function $\rho_+(\bm{p},\omega)$ 
for $T/m=0.4, 0.8, 1$ and $1.2$.}
\label{fig:spcP}
\end{center}
\vspace{-5mm}
\end{figure}

In Fig.~\ref{fig:spcP},
we plot the quark spectral function
for $T/m = 0.4 , 0.8, 1.0$ and $1.2$.
It is seen that there appears a three-peak structure
at intermediate temperatures, as found in the previous section.
This is quite natural, and it would be expected from the
behavior of the self-energy.
A clear three-peak structure is formed at lower values of $T$
than in the scalar boson case.
This is because 
the term $( \omega^2 +2m^2 ) / m^2$ in Eq.~(\ref{eq:ImSigP_p0}), 
which gives approximately a factor of $2$ for $\omega \lesssim m $,
tends to enhance the coupling constant $g$.
As discussed in \S\ref{subsec:g},
the temperature at which the clear three-peak structure appears
decreases for larger $g$.

\subsection{Quark spectrum at high temperatures}

\begin{figure}[t]
\begin{center}
\begin{tabular}{cc}
\hspace{-1.8cm} $T/m=5$ & \hspace{-2.3cm} $T/m=10$ \\
\hspace{-.7cm}
\includegraphics[width=220pt]{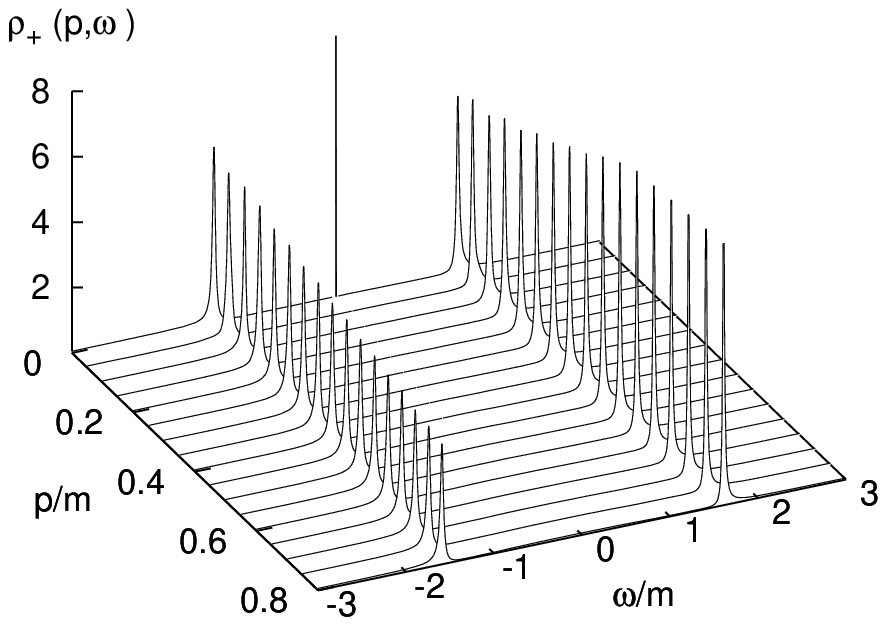} & \hspace{-1.3cm}
\includegraphics[width=220pt]{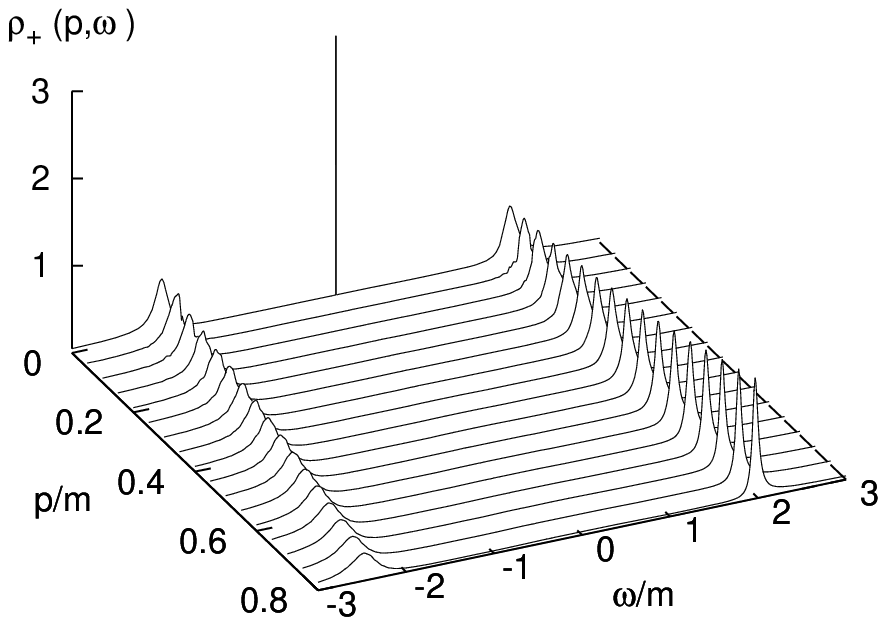}
\end{tabular}
\caption{The quark spectral function $\rho_+(\bm{p},\omega)$ 
for $T/m=5$ and $10$.
}
\label{fig:spcP2}
\end{center}
\end{figure}

Next, we plot the quark spectral function 
at higher temperatures, $T/m = 5$ and $10$,
in Fig.~\ref{fig:spcP2}.
We find that the energy at the quasi-particle peak
converges to a value near $\omega/m = 2.5 $ 
as $T/m$ increases for $\bm{p}=0$.
It is also noteworthy that the two peaks with finite 
thermal mass become obscure as $T$ increases.

\begin{figure}[t]
\begin{center}
\includegraphics[width=0.49\textwidth]{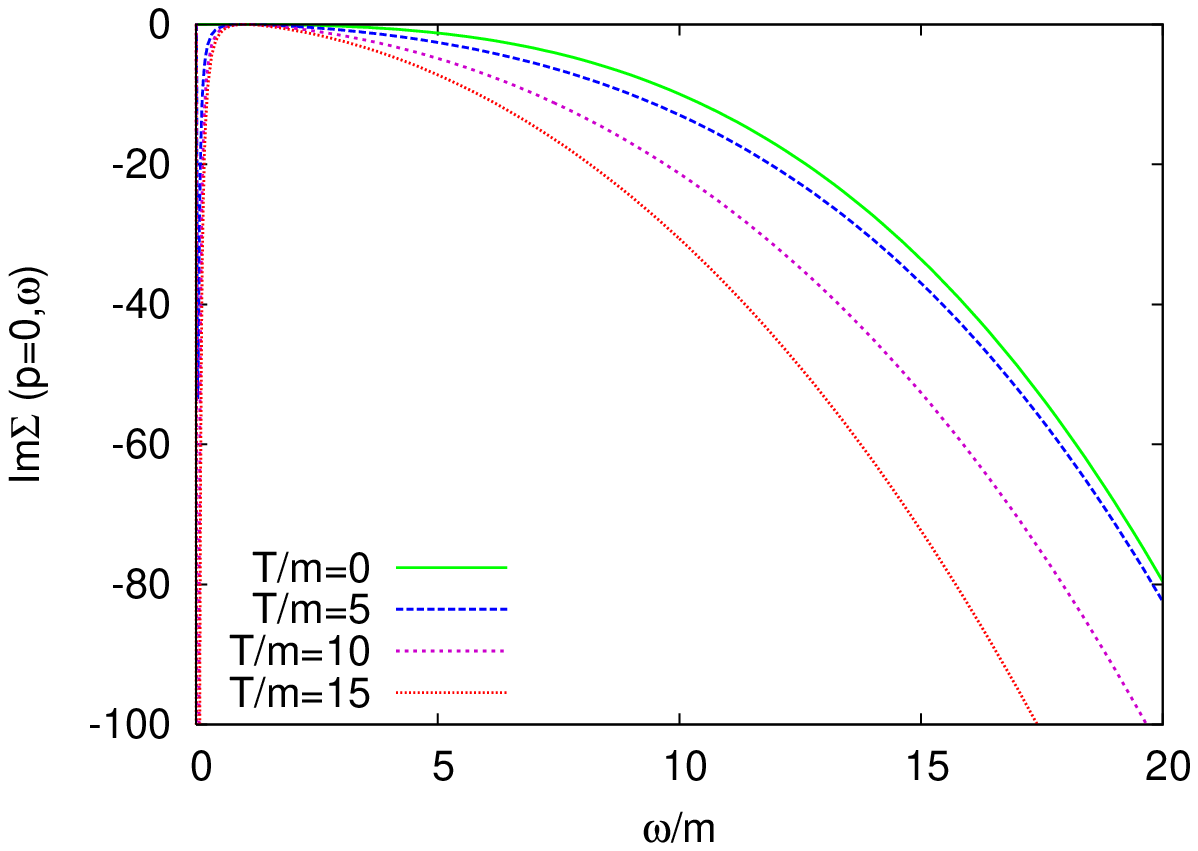}
\includegraphics[width=0.49\textwidth]{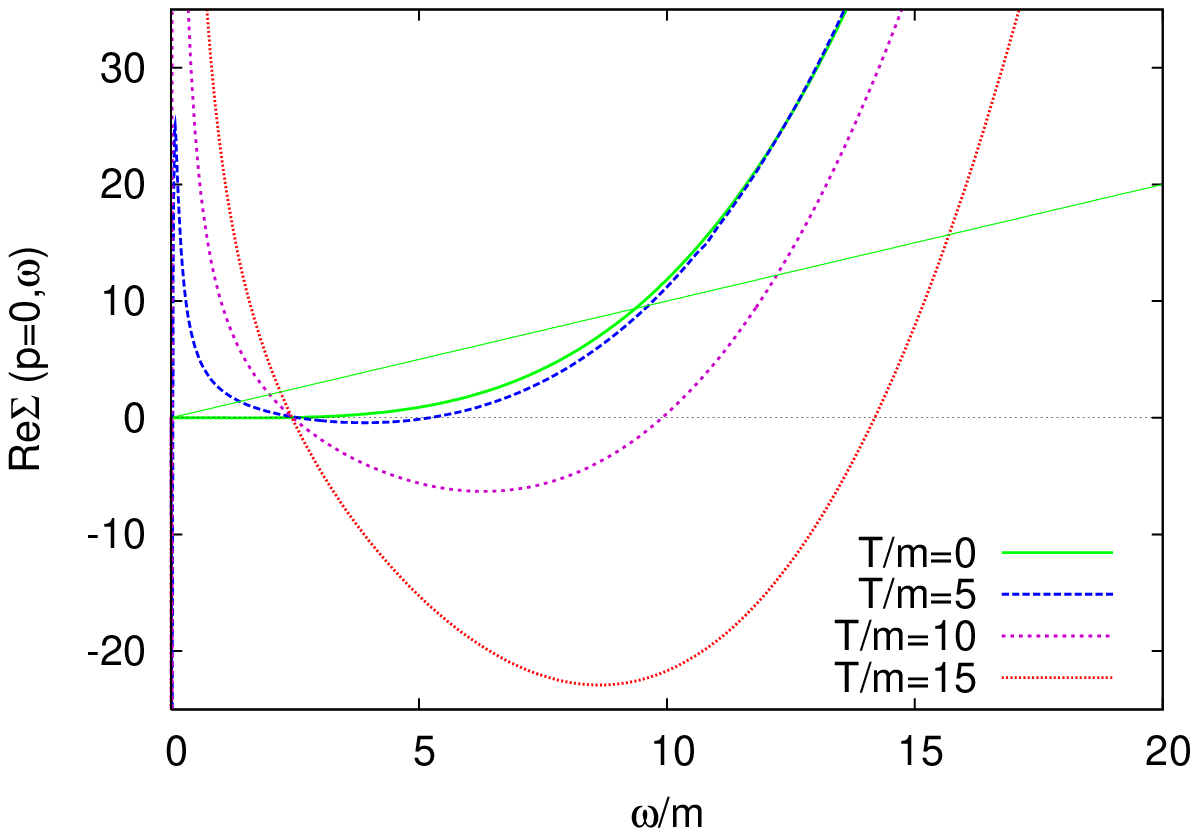}
\caption{
The imaginary and real parts of the self-energy, $\Sigma_+ (\bm{p},\omega)$,
at zero momentum for $T/m=5,10$ and $15$.}
\label{fig:ImSigmaP2}
\end{center}
\end{figure}

In order to understand this behavior of 
$ \rho_+ ( \bm{p},\omega )$ at higher $T$,
in Fig.~\ref{fig:ImSigmaP2} 
we plot the self-energy for $T/m=0, 5, 10$, and $15$.
In the right figure, we plot the line 
${\rm Re}\Sigma_+ = \omega$ in order to see 
the quasi-dispersion relation.
At $T=0$, there exist three quasi-dispersion relations
at $\omega=0$ and near $\omega = \pm 10m$.
The latter two solutions are
 unphysical.\footnote{These quasi-dispersion relations do not form a peak
in the quark spectral function, because
$|{\rm Im}\Sigma^R|$ takes a large value there.
We have confirmed that they also exist for very large values
of $\omega$ in the model with a scalar boson
in \S\ref{sec:yukawa}.}
In Fig.~\ref{fig:ImSigmaP2}, 
it is also noteworthy that
${\rm Re}\Sigma_+$ changes sign near $\omega=2.5m$
for higher $T$.
In fact, we can analytically prove that the zero of 
${\rm Re}\Sigma_+$ approaches $\omega = \sqrt6 m$
in the high $T$ limit (see Appendix~\ref{app:vector_asym}).
Due to this property, the solution of the quasi-dispersion relation
corresponding to the two peaks in Fig.~\ref{fig:spcP}
approaches $|\omega| = \sqrt6 m$ at $\bm{p}=0$ in the high $T$ limit.

\section{Summary and concluding remarks} \label{sec:conc}

We have investigated the quasi-particle picture of a fermion,
which we call a ``quark" in this paper, employing Yukawa models 
composed of a massless quark and  a massive boson
at finite temperature ($T$).
We have 
considered four types of massive bosons, i.e.,
scalar, pseudoscalar, vector and  axial-vector bosons, 
with a mass $m$,
though there is no difference in the quark spectra
for the scalar (vector)
and pseudoscalar (axial-vector) bosons,
because the quark is massless.
The Proca formalism was employed for the vector boson.
For both models, the quark self-energy was computed up to
one-loop order, and the spectral function and quasi-dispersion
relation of the quark were studied.
We have found that
the quark spectral function at low frequency and low momentum
is significantly different from the free quark spectrum, and
it has a three-peak structure at intermediate values of $T$, i.e., 
for $T/m\sim 1$.
We have also found that the number of branches of the
quasi-dispersion relation does not necessarily 
coincide with that of the peaks in the fermion 
spectral function when the imaginary part of the quark self-energy
is large. 
We have shown that the two of the three peaks have
finite thermal masses and approach the normal quasi-quark and plasmino 
excitations in the HTL approximation in the high $T$ limit, 
while the strength of the peak at the origin becomes weaker and disappears
in this limit.
In the Yukawa model with a vector boson, 
we have found that the behavior of 
the spectral function is the same as that
in the Yukawa model with a scalar boson up to $T\approx m$.
In the high $T$ limit,
the spectrum in the Proca formalism does not approach that
in the HTL approximation.

We have discussed the fact that the three-peak structure originates from
the Landau damping, that is, the scattering processes
of a quark and an antiquark hole of a thermally excited antiquark,
and of an antiquark and a quark hole of a thermally excited quark. 
They cause the formation of energy gaps
 in the quark spectrum,
owing to the level mixing between the quark (antiquark) and the hole of 
the thermally excited antiquarks (quarks).
This leads to a level repulsion or gap in the resultant physical spectrum.
These mixings can be understood in terms of the resonant scattering
\cite{Janko1997,Kitazawa:2005pp} of the quasi-quarks off the bosons.
In particular, owing to the mass of the boson, the level repulsion
due to the resonant scattering occurs \textit{twice}
at different points in the energy-momentum plane, 
leading to the three-peak structure of the spectral function.
This contrasts with the case of the quark spectrum coupled to a 
massless boson,
such as a gauge boson, in which the resonant scattering occurs only at
the origin in the energy-momentum plane and then leads to only two peaks,
as in the spectrum in the HTL approximation.

Since there may exist  bosonic 
excitations (even apart from  gluons) coupled to quarks
in the QGP phase\cite{Hatsuda:1985eb, Shuryak:2004cy, Brown:2003km},
it is plausible that 
the quark spectrum would have a three-peak structure 
in the QGP phase if the boson has a ``mass'' of the order of $T$;
if there exist several bosonic excitations with ``masses'' close to
$T$, the quark spectral function may become a superposition of
three-peak structures with various heights.
Although the quark spectral function in QCD is 
a gauge-dependent quantity, the complex pole of the quark 
Green function is gauge invariant.
For this reason, we conjecture that the peak structure of the quark
spectral function reflects the complex pole of the Green function
and will be seen in gauge theories, for example, QCD.

One interesting example of bosonic excitations
which may affect the quasi-particle picture of the quarks in the
QGP phase is the soft modes associated 
with the chiral transition\cite{Hatsuda:1985eb}.
The chiral soft mode is unique, because its effective mass
 $m^*_\sigma(T)$ may become as small as a few hundred MeV,
 and thus it can eventually be of
the same order as $T$ near $T_c$ of the chiral transition
if the phase transition is second order or weakly first order.
In fact, it has been  shown\cite{Kitazawa:2005mp} 
that the chiral soft modes give rise to a three-peak structure 
in the quark spectrum just above $T_c$:
In Ref.~\citen{Kitazawa:2005mp},  the Nambu-Jona-Lasinio model is employed
 as a chiral model, and the quark spectral function 
is calculated incorporating the effect of 
the dynamically generated soft modes into the quark propagator.
It is noteworthy  that although the soft mode obtained in the dynamical model
employed in Ref.~\citen{Kitazawa:2005mp}
is a composite particle with a finite width and 
thus not an elementary boson,
a clear three-peak structure is realized in the quark spectral function.

It would be interesting to investigate the effect in the change of
the quark spectrum on various observables in the QGP phase.
In particular, it would be intriguing to explore the 
dilepton production rate;
because the quasiparticle peaks in the three-peak structure
have low sound velocities,
there can appear a sharp peak in the dilepton production rate
owing to a mechanism similar to that causing the van Hove 
singularities\cite{Braaten:1990wp}.
It would also be an interesting to consider the properties
of mesonic excitations in the QGP phase 
composed of quarks having the spectrum obtained in this work.

We remark that 
a lattice QCD simulation of the quark spectrum 
would be, of course, preferable, 
although this would require the performance of several tasks, such as
the gauge fixing,  carrying out the chiral extrapolation
and using dynamical (chiral) fermions to elucidate the effects of the 
chiral soft modes.

Contributions of higher order terms such as more than 
one-loop diagrams,  which are not taken into account in the 
present, may affect the peak structure of the spectrum. 
In condensed matter physics, some corrections based on self-consistency 
conditions are proposed\cite{Yanase} and their possible effects on 
the quark spectrum 
near the color superconducting phase transition are commented in 
Ref. \citen{Kitazawa:2004cs2}.

Because the present investigation is based on simple but generic
Yukawa models with a massive scalar and vector boson,
the results obtained here should apply to various systems
at finite $T$ composed of a light fermion and a massive boson,
in addition to QCD matter in the QGP phase. 
One such system is the excitation spectra of neutrinos 
coupled to weak bosons at high $T$
near the masses of the weak bosons\cite{Boyanovsky:2005hk}.
The investigation of a such system would be carried out 
exclusively in terms of 
the neutrino quasi-dispersion relations.

In this paper,  we have considered only massless quarks.
A finite quark mass should affect the
formation of the three-peak structure in the quark spectral
function. Thus, the
incorporation of the quark mass effect is important for describing the 
chiral transition precisely,
because the order of
the transition, and hence,
the quark spectrum near the critical point are sensitive 
to the quark mass. 
A finite quark mass also leads to a difference
between the scalar (vector) and pseudoscalar (axial-vector) boson 
cases, and may
suppress the three-peak structure, 
because it does not appear in Ref. \citen{Baym:1992eu},
where a massive quark spectrum coupled with a massless boson 
is studied at finite $T$. 
Detailed study of such a effect is now under way
and will be reported elsewhere\cite{MKKN}.
The finite chemical potential may also alter the three-peak structure
because antiquarks in medium are fewer in this case, and thus, 
mixing between the quark and the antiquark-hole will be suppressed. 
Such a study is also now under investigation\cite{KKN}.

\section*{Acknowledgements}

M. K. is grateful to Y.~Hidaka and R.~D.~Pisarski for useful discussions.
He also acknowledges L.~McLerran and F.~Karsch for their interest 
in this work. T. K. acknowledges B.~M\"{u}ller for valuable comments.
Y. N. thanks M. Harada for useful comments.
M. K. is supported by a Special Postdoctoral Research Program of RIKEN.
T. K. is supported by a Grant-in-Aid
for Scientific Research by Monbu-Kagakusyo of Japan
(No. 17540250).
Y.~N. is supported by the 21st Century COE Program at Nagoya University
and a Grand-in-Aid for Scientific Research by Monbu-Kagakusyo 
of Japan (No. 18740140).
This work is supported by the Grant-in-Aid for the 21st Century COE 
``Center for Diversity and Universality in Physics" of Kyoto
University.

\appendix


\section{Quark Self-Energy in the Yukawa Model with Scalar Coupling} 
\label{app:sigma}

In this appendix, 
we summarize the calculation of the quark self-energy 
in the model with the scalar boson given in Eq.~(\ref{eq:tildeSigma}),
\begin{eqnarray}
  \tilde\Sigma ( \bm{p},i\omega_m ) = 
  -g^2 T \sum_n \int \frac{ d^3\bm{k} }{ (2\pi)^3 }
  {\cal G}_0 ( \bm{k},i\omega_n )
  {\cal D}( \bm{p}-\bm{k} ,i\omega_m-i\omega_n ),
  \label{eq:app:tildeSigma}
\end{eqnarray}
where 
\begin{eqnarray}
 {\cal G}_0 ( \bm{k},i\omega_n )
 &=& \frac{1}{i\omega_n \gamma^0 - \bm{k}\cdot\bm{\gamma}}
 = \sum_s \Lambda_s (\bm{k}) \gamma^0 \frac{1}{i\omega_n - s E_f},
  \label{eq:freeq}\\
 {\cal D} ( \bm{k},i\nu_n ) 
  &=& \frac{1}{ (i\nu_n)^2 - \bm{k}^2 - m^2 }
  = \sum_t t \frac{1}{2E_b( i\nu_n - t E_b )}
  \label{eq:freeb}
\end{eqnarray}
are the free propagators of  the quark and the scalar boson, 
respectively. Here,
$\Lambda_\pm (\bm{k}) = ( 1 \pm \gamma^0 \bm{\gamma}\cdot\hat{\bm k} ) /2 $
are the projection operators onto positive and negative energies, respectively.
Substituting Eqs.~(\ref{eq:freeq}) and (\ref{eq:freeb}) 
into Eq.~(\ref{eq:app:tildeSigma}), we have
\begin{eqnarray}
  \tilde\Sigma ( \bm{p},i\omega_m )
  &=& -g^2 \sum_{ s,t=\pm }
  T\sum_n \int \frac{ d^3 \bm{k} }{ (2\pi)^3 }
  \frac{ t \Lambda_s (\bm{k}) \gamma^0 }{ 2 E_b }
  \frac1{ i\omega_n - sE_f }.
  \frac1{ i\omega_m - i\omega_n - tE_b },\notag \\
  \label{eq:app:tildeSigma2}
\end{eqnarray}
with $E_f= |\bm{k}|$ and $ E_b=\sqrt{ (\bm{p}-\bm{k})^2 + m^2 } $.
Carrying out the summation over the Matsubara modes $n$ and applying
the analytic continuation $i\omega_m \to \omega+i\eta$ 
in Eq.~(\ref{eq:app:tildeSigma2}), we obtain
the retarded self-energy 
and its imaginary part:
\begin{eqnarray}
\Sigma^R ( \bm{p},\omega )
&=& -g^2 \sum_{ s,t=\pm }
\int \frac{ d^3 \bm{k} }{ (2\pi)^3 }
\frac{ t \Lambda_s (\bm{k}) \gamma^0 }{ 2 E_b }
\frac{  f(sE_f) + n(-tE_b) }{ \omega - sE_f - tE_b + i\eta },
\label{eq:app:Sigma^R}
\\
{\rm Im} \Sigma^R ( \bm{p},\omega )
&=& \pi g^2 \sum_{ s,t=\pm }
\int \frac{ d^3 \bm{k} }{ (2\pi)^3 }
\frac{ t \Lambda_s (\bm{k}) \gamma^0 }{ 2 E_b }
\left\{  f(sE_f) + n(-tE_b)  \right\}
\delta( \omega - sE_f - tE_b ). \notag \\
\label{eq:app:ImSigma^R}
\end{eqnarray}
Equations (\ref{eq:sigmar}) and (\ref{eq:imsig}) are obtained
by carrying out the summations over $s$ and $t$ in 
Eqs.~(\ref{eq:app:Sigma^R})
and (\ref{eq:app:ImSigma^R}), respectively.

We next decompose the self-energy with the projection operators into
$\Sigma^R( \bm{p},\omega )
= \Sigma_+( \bm{p},\omega ) \Lambda_+(\bm{p})\gamma^0
+ \Sigma_-( \bm{p},\omega ) \Lambda_-(\bm{p})\gamma^0
$.
The imaginary part of $\Sigma_+ ( \bm{p},\omega )$ is
calculated in the following way:
\begin{eqnarray}
{\rm Im}\Sigma_+( \bm{p},\omega ) 
&=& 
\frac12 {\rm Tr} [{\rm Im}\Sigma^R ( \bm{p},\omega ) 
\Lambda_+ (\bm{p}) \gamma^0 ]
\nonumber \\
&=&
\pi g^2 \sum_{ s,t=\pm }
\int \frac{ d^3 \bm{k} }{ (2\pi)^3 }
\frac{ 1 - s \hat{\bm p}\cdot \hat{\bm k} }{ 4 t E_b }
\left\{  f(sE_f) + n(-tE_b)  \right\}
\delta( \omega - s E_f - t E_b )
\label{eq:app:ImSigma^R_+a0} \notag \\
& & \\
&=&
-\frac{ g^2 }{ 32\pi \bm{p}^2 }
\sum_{ s,t=\pm } st
\int d E_f \int_{e_-}^{e_+} dE_b
\left[ ( |\bm{p}| - sE_f )^2 - E_b^2 + m^2 \right] \notag \\
& & \qquad \times\left\{  f(sE_f) + n(-tE_b)  \right\}
\delta( \omega - s E_f - t E_b )
\nonumber \\
&=&
-\frac{ g^2 }{ 32\pi \bm{p}^2 }
\sum_{ s,t=\pm } st
\int d E_f 
\left[ ( |\bm{p}|+\omega - 2E_f )( |\bm{p}| - \omega ) + m^2 \right] \notag \\
& & \qquad \times \left\{  f(sE_f) + n( sE_f-\omega )  \right\}
\int_{e_-}^{e_+} dE_b
\delta( \omega - s E_f - t E_b ),
\label{eq:app:ImSigma^R_+a}
\end{eqnarray}
where in the third equality we have used the relations
\begin{eqnarray}
\hat{\bm p}\cdot\hat{\bm k} 
&=& \frac{ \bm{p}^2 + E_f^2 - E_b^2 + m^2 }{ 2 |\bm{p}| E_f },
\label{eq:app:pk} \\
\int \frac{ d^3 \bm{k} }{ (2\pi)^3 }
&=& \frac1{ 4\pi^2 |\bm{p}| } \int dE_f E_f \int_{e_-}^{e_+} dE_b E_b,
\label{eq:app:integ}
\end{eqnarray}
with $e_\pm = \sqrt{ ( |\bm{p}|\pm E_f )^2 + m^2 }$.

In the time-like region, 
one of the four terms 
in Eq.~(\ref{eq:app:ImSigma^R_+a}) takes a finite value 
(see Fig.~\ref{fig:imsig}).
After some algebra,
it is found that all these terms in the time-like region 
are reduced to the following simple form:
\begin{eqnarray}
{\rm Im}\Sigma_+( \bm{p},\omega ) 
&=& 
- \frac{ g^2 }{ 32\pi \bm{p}^2 }
\int_{E_f^+}^{E_f^-} dE_f
\left[ ( |\bm{p}|+\omega - 2E_f )( |\bm{p}| - \omega ) + m^2 \right] \notag \\
& & \qquad \times \left\{  f(sE_f) + n( sE_f-\omega )  \right\},
\label{eq:app:ImSigma^R_+b}
\end{eqnarray}
with
\begin{eqnarray}
E_f^\pm = \frac{ \omega^2 - \bm{p}^2 - m^2  }{ 2( \omega \pm |\bm{p}| ) }.
\end{eqnarray}
In the space-like region, the two terms corresponding to (II) and (III)
in Fig.~\ref{fig:feynykw}
take finite values, and Eq.~(\ref{eq:app:ImSigma^R_+a})
yields
\begin{eqnarray}
&&\hspace*{-5mm}{\rm Im}\Sigma_+( \bm{p},\omega )\nonumber \\
&=& 
- \frac{ g^2 }{ 32\pi \bm{p}^2 }
\left[ \int_{E_f^+}^{-\infty} dE_f - \int_{\infty}^{E_f^-} dE_f  \right]
\left[ ( |\bm{p}|+\omega - 2E_f )( |\bm{p}| - \omega ) + m^2 \right] \notag \\
& & \qquad \times \left\{  f(E_f) + n( E_f-\omega )  \right\}
\nonumber \\
&=&
- \frac{ g^2 }{ 32\pi \bm{p}^2 }
\left[ \int_{E_f^+}^{E_f^-} dE_f
- \int_{-\infty}^{\infty} dE_f \right]
\left[ ( |\bm{p}|+\omega - 2E_f )( |\bm{p}| - \omega ) + m^2 \right] \notag \\
& & \qquad \times \left\{  f(E_f) + n( E_f-\omega )  \right\}
\nonumber \\
&=&
- \frac{ g^2 }{ 32\pi \bm{p}^2 }
\int_{E_f^+}^{E_f^-} dE_f
\left[ ( |\bm{p}|+\omega - 2E_f )( |\bm{p}| - \omega ) + m^2 \right]
\left\{  f(E_f) + n( E_f-\omega )  \right\}
\nonumber \\
&&
- \frac{ \pi g^2 T^2 }{ 32 } \frac{ |\bm{p}|-\omega }{ \bm{p}^2 }
- \frac{ g^2 }{ 32\pi } 
\frac{ \omega [ |\bm{p}| ( |\bm{p}|-\omega ) + m^2 ] }{ \bm{p}^2},
\label{eq:app:ImSigma^R_+c}
\end{eqnarray}
where  we have used the integral formulae
\begin{eqnarray}
\int_{-\infty}^{\infty} d \epsilon 
\left[ f(\epsilon) + n( \epsilon-\omega ) \right] 
= -\omega,
\\
\int_{-\infty}^{\infty} d \epsilon \epsilon
\left[ f(\epsilon) - n( \epsilon-\omega ) \right] 
= \frac{ \pi^2 T^2 }2 + \frac{ \omega^2 }2.
\end{eqnarray}
Equations~(\ref{eq:app:ImSigma^R_+b}) and (\ref{eq:app:ImSigma^R_+c})
are combined to give
\begin{eqnarray}
&&\hspace{-5mm}{\rm Im}\Sigma_+( \bm{p},\omega ) \nonumber\\
&=& 
- \frac{ g^2 }{ 32\pi \bm{p}^2 }
\int_{E_f^+}^{E_f^-} dE_f
\left[ ( |\bm{p}|+\omega - 2E_f )( |\bm{p}| - \omega ) + m^2 \right]
\left\{  f(E_f) + n( E_f-\omega )  \right\}
\nonumber \\
&&
- \left\{
\frac{ \pi g^2 T^2 }{ 32 } \frac{ |\bm{p}|-\omega }{ \bm{p}^2 }
+ \frac{ g^2 }{ 32\pi } 
\frac{ \omega [ |\bm{p}| ( |\bm{p}|-\omega ) + m^2 ] }{ \bm{p}^2}
\right\} \theta( \bm{p}^2 -\omega^2 ).
\label{eq:app:ImSigma^R_+}
\end{eqnarray}

For $\bm{p}=\bm{0}$, 
from Eq.~(\ref{eq:app:ImSigma^R_+a0})
we obtain
\begin{eqnarray}
{\rm Im}\Sigma_\pm ( \bm{0},\omega )
= - \frac{g^2}{64\pi} \frac{ ( \omega^2 - m^2 )^2 }{ \omega^3 }
\left( \coth \frac{ \omega^2 + m^2 }{ 4T \omega }
     + \tanh \frac{ \omega^2 - m^2 }{ 4T \omega } \right).
\end{eqnarray}

The $T$-independent part of ${\rm Im}\Sigma_+(\bm{p},\omega)$
is given by the $T=0$ limit of Eq.~(\ref{eq:app:ImSigma^R_+}).
In this case,
the integral in Eq.~(\ref{eq:app:ImSigma^R_+}) can be performed
analytically, and we obtain
\begin{eqnarray}
{\rm Im}\Sigma_+( \bm{p},\omega )_{T=0} 
&=& 
- \frac{g^2}{32\pi} (\omega-|\bm{p}| ) 
\frac{ ( P^2 - m^2 )^2 }{ P^4 }
{\rm sgn}(\omega) \theta( P^2 - m^2 ),
\label{eq:app:ImSigma^R_+d}
\end{eqnarray}
with $P^2 = \omega^2 - \bm{p}^2$.

The $T$-dependent part, 
${\rm Im}\Sigma^R_+ ( \bm{p},\omega )_{T\ne0}$,
is obtained through the replacement
of the distribution function in the curly brackets
in Eq.~(\ref{eq:app:ImSigma^R_+a}) as
\begin{eqnarray}
f(E_f)+n( \omega - E_f )
\to n( \omega - E_f ) + f(E_f) 
+ \epsilon(\omega) \theta( P^2 - m^2 ).
\label{eq:app:Tne0}
\end{eqnarray}

\section{Quark Self-Energy in the Yukawa Model with Vector Coupling} 
\label{app:vector}

The self-energy of the quark in a Yukawa model with a massive vector boson
is given by Eq.~(\ref{eq:SigmaVImag}) as
\begin{eqnarray}
\tilde\Sigma ( \bm{p} , i\omega_m )
= -g^2 T \sum_n \int \frac{ d^3 \bm{k} }{ (2\pi)^3 }
\gamma^\mu {\cal G}_0( \bm{k},i\omega_n ) \gamma^\nu 
{\cal D}_{\mu\nu} ( \bm{p}-\bm{k} , i\omega_m-i\omega_n ),
\label{eq:app:SigmaV}
\end{eqnarray}
with the Proca propagator ${\cal D}_{\mu\nu}$ defined in
Eq.~(\ref{eq:Proca}).
Carrying out the summation over the Matsubara modes $n$ and 
applying the analytic continuation
$i\omega_m\to \omega+i\eta$,
the retarded self-energy 
$\Sigma^R( \bm{p},\omega ) $ is obtained as Eq.~(\ref{eq:SigmaVector}), and 
the imaginary part is given by
\begin{eqnarray}
{\rm Im}\Sigma^R ( \bm{p},\omega )
&=&
-\pi g^2 \sum_{s,t=\pm} t \int \frac{ d^3 \bm{k} }{ (2\pi)^3 }
\frac { \gamma^\mu \Lambda_s (\bm{k}) \gamma^0 \gamma^\nu }{2E_b} 
\left( g_{\mu\nu} - \frac{ q_\mu q_\nu }{m^2} \right)
\nonumber \\
&& \times
\left\{ f( sE_f ) + n( -tE_b ) \right\}
\delta ( \omega - sE_f - tE_b ),
\label{eq:app:ImSigmaV}
\end{eqnarray}
with $E_f = |\bm{k}|$, $E_b = \sqrt{ ( \bm{p}-\bm{k} )^2 + m^2 }$
and $q_\mu = ( tE_b , \bm{p} - \bm{k} )$.
The quark sector of Eq.~(\ref{eq:app:ImSigmaV}) gives
\begin{eqnarray}
{\rm Im}\Sigma_+ ( \bm{p},\omega )
&=&
{\rm Tr} \left[ {\rm Im} \Sigma^R ( \bm{p},\omega )
\Lambda_+(\bm{p}) \gamma^0 \right]
\nonumber \\
&=&
\pi g^2 
\sum_{s,t=\pm} t \int \frac{ d^3 \bm{k} }{ (2\pi)^3 }
\frac{t}{2E_b} \left[ \frac{ (tE_b - s\hat{\bm k}\cdot \bm{q} )
( tE_b - \hat{\bm p}\cdot \bm{q} ) }{ m^2 }
- \frac{ 1-s \hat{\bm k} \cdot \hat{\bm p} }{2} \right]
\nonumber \\
&& \times
\left\{ f( sE_f ) + n( -tE_b ) \right\}
\delta ( \omega - sE_f - tE_b )
\nonumber \\
&=&
\frac{ g^2 }{ 32\pi \bm{p}^2 m^2 }
\sum_{s,t=\pm} st 
\int dE_f 
[ ( \bm{p}^2 - \omega^2 + m^2 )\left( (|\bm{p}|-\omega)^2 
- 2m^2 \right) \notag\\
& & \qquad + 2sE_f ( \bm{p}^2 - \omega^2 + 2m^2 )( |\bm{p}|-\omega ) ]
\int_{e^-}^{e^+} dE_b 
\delta ( \omega - sE_f - tE_b ) \notag \\
&=&
- \frac{ g^2 }{ 32\pi \bm{p}^2 m^2 }
\int_{E_f^+}^{E_f^-} d\epsilon
[ - ( P^2 - m^2 )\{ ( |\bm{p}|-\omega )^2 - 2m^2 \} \notag \\
& & \qquad + 2\epsilon ( \omega-|\bm{p}| )( P^2 - 2m^2 ) ]
\left\{ f(\epsilon) + n( \epsilon-\omega ) \right\}
\nonumber \\
&&- \frac{ g^2 }{ 32\pi \bm{p}^2 m^2 } \theta( - P^2 )
\bigg\{ ( P^2 - 2m^2 )(\omega-|\bm{p}|) \pi^2 T^2 \notag \\
& & \qquad +\omega [ 2m^4 - P^2 \{ |\bm{p}|(\omega-|\bm{p}|) + m^2 \} ] 
\bigg\}.
\label{eq:app:imsigv}
\end{eqnarray}
where in the third equality we have used Eqs.~(\ref{eq:app:pk}),
(\ref{eq:app:integ}) and
\begin{eqnarray}
\hat{\bm k}\cdot \bm{q}
= \frac{ \bm{p}^2 - E_f^2 - E_b^2 + m^2 }{ 2E_f },
\quad
\hat{\bm p}\cdot \bm{q}
= \frac{ \bm{p}^2 - E_f^2 + E_b^2 - m^2 }{ 2|\bm{p}| }.
\end{eqnarray}

For $\bm{p}=\bm{0}$, we have
\begin{eqnarray}
{\rm Im}\Sigma_\pm ( \bm{0},\omega )
= - \frac{g^2}{64\pi} \omega 
\frac{ \omega^2 + 2m^2 }{ m^2 }
\frac{ ( \omega^2 - m^2 )^2 }{ \omega^4 }
\bigg( \coth \frac{ \omega^2 + m^2 }{ 4T \omega } 
+ \tanh \frac{ \omega^2 - m^2 }{ 4T \omega } \bigg).\notag \\
\end{eqnarray}

For the $T$-independent part, which is obtained by taking the 
$T\to 0$ limit in Eq.~(\ref{eq:app:imsigv}),
we can perform the integral in Eq.~(\ref{eq:app:imsigv}) analytically,
and we obtain
\begin{eqnarray}
{\rm Im}\Sigma_+( \bm{p},\omega )_{T=0} 
&=& 
- \frac{g^2}{32\pi} (\omega-|\bm{p}| ) 
\frac{ P^2 + 2m^2 }{ P^2 }
\frac{ ( P^2 - m^2 )^2 }{ P^4 }
{\rm sgn}(\omega) \theta( P^2 - m^2 ),
\label{eq:app:ImSigmaT=0}\qquad
\end{eqnarray}
with $P^2 = \omega^2 - \bm{p}^2$.

\section{Renormalization of the $T$-Independent Part}
\label{app:T=0}

In this appendix, we perform the renormalization of the $T$-independent part of
the self-energy, $\Sigma ( \bm{p},\omega )_{T=0}$, in the models studied
in \S\S\ref{sec:yukawa} and \ref{sec:vector}.
Because the imaginary part of $\Sigma^R ( \bm{p},\omega )_{T=0}$
in both models is free from ultraviolet divergence, as shown in 
Appendices \ref{app:sigma} and \ref{app:vector}, renormalization
is needed for the real part only.
The $T$-independent part is Lorentz covariant.
To calculate $\Sigma ( \bm{p},\omega )_{T=0}$ by using
this symmetry explicitly, we employ the
self-energy in the Feynman propagator $\Sigma^F$
instead of the retarded one, $\Sigma^R$. 
Once the renormalized self-energy in the Feynman propagator is derived,
the corresponding retarded one is readily obtained.
In the following, 
we omit the subscript of the self-energy, writing $\Sigma^F_{T=0}$ as
$\Sigma^F$.

Because of the Lorentz invariance, the self-energy has the form
\begin{eqnarray}
\Sigma^F( \bm{p},\omega )
= p\sla \Sigma^F_1( P^2 ) + \Sigma^F_2( P^2 ),
\end{eqnarray}
with $p^\mu = ( \omega , \bm{p} )$.
For the self-energy used in \S\ref{sec:yukawa} and \S\ref{sec:vector},
$\Sigma_2(P^2)$ vanishes.

We regularize the real part of the self-energy by employing
the subtracted dispersion relation.
This is much more convenient than other regularization procedures,
e.g., the Pauli-Villars and dimensional regularizations.
(We have confirmed that these regularizations give 
the same result in our models.)
The subtracted dispersion relation is written
\begin{eqnarray}
  {\rm Re}\Sigma_1^F(P^2)=\sum_{l=0}^{n-1} \frac{(P^2-\alpha)^l}{l!}
  c_l +
  \frac{(P^2-\alpha)^n}{\pi} \textrm{P}\int_{s_{\rm thr}}^\infty
  ds \frac{{\rm Im}\Sigma_1^F(s)}{(s-P^2)(s-\alpha)^n},
  \label{eq:app:SDR}
\end{eqnarray}
where $\alpha$ denotes the subtraction point or the
renormalization point, and $s_{\rm thr}$ denotes the threshold,
which is given by $s_{\rm thr}=m^2$ in our case.
The parameters $c_l$ are the subtraction constants, which are to be
determined by the renormalization condition.
Equation~(\ref{eq:app:SDR}) is valid for $\alpha<s$.

\subsection{Renormalization in the Yukawa model with scalar coupling}

From Appendix~\ref{app:sigma},
the imaginary part of $\Sigma_1^F(P^2)$ reads
\begin{eqnarray}
{\rm Im}\Sigma^F_1( P^2 )
= - \frac{ g^2 }{ 32\pi }
\frac{ ( P^2-m^2 )^2 }{ P^4 } \theta( P^2 - m^2 ).
\label{eq:app:ImSigma_1}
\end{eqnarray}
For the real part of $\Sigma^F_1$,
we use the subtracted dispersion relation given in Eq.~(\ref{eq:app:SDR})
for $n=1$ to regularize the ultraviolet divergence.
We impose the on-shell renormalization condition,
$\Sigma_1^F(P^2=\alpha)=0$ and $\alpha=0$.
From Eq.~(\ref{eq:app:ImSigma_1}), we have ${\rm Im}\Sigma_1^F(P^2=0)=0$.
The renormalized real part is then given by
\begin{eqnarray}
{\rm Re}\Sigma^F_1(P^2)
&=&
\frac{ P^2 }\pi {\rm P} \int_{m^2}^\infty ds 
\frac{ {\rm Im}\Sigma^F_1(s) }{ s ( s-P^2 ) }
\nonumber \\
&=&
\frac{ g^2 }{ 32\pi^2 } \left\{
\frac{ ( P^2-m^2 )^2 }{ P^4 }
\log \left| \frac{ P^2 - m^2 }{ m^2 } \right|
- \frac32 + 2 \frac{ m^2 }{ P^2 }
\right\}.
\end{eqnarray}
The retarded self-energy, $\Sigma^R( \bm{p},\omega )$, is given by 
${\rm Re}\Sigma^R( \bm{p},\omega ) = {\rm Re}\Sigma^F( P^2 )$
and \linebreak ${\rm Im}\Sigma^R( \bm{p},\omega ) 
= \epsilon(\omega){\rm Im}\Sigma^F( P^2 )$:
\begin{eqnarray}
\Sigma^R( \bm{p},\omega )
&=&
\frac{ g^2 }{ 32\pi^2 } p\sla \left\{
\frac{ ( P^2-m^2 )^2 }{ P^4 }
\log \left| \frac{ P^2 - m^2 }{ m^2 } \right|
- \frac32 + 2 \frac{ m^2 }{ P^2 }
\right\} \notag \\
& & -i \frac{ g^2 }{ 32\pi } p\sla \frac{ ( P^2-m^2 )^2 }{ P^4 } 
\epsilon(\omega) \theta( P^2 - m^2 ).
\end{eqnarray}

\subsection{Renormalization in the Yukawa model with vector coupling}

From Eq.~(\ref{eq:app:ImSigmaT=0}),
the imaginary part of the self-energy $\Sigma^F_1$ reads
\begin{eqnarray}
{\rm Im}\Sigma^F_1(P^2)
&=&
- \frac{ g^2 }{ 32\pi } 
\frac{ P^2 + 2m^2 }{ m^2 }
\frac{ ( P^2-m^2 )^2 }{ P^4 }
\theta ( P^2 - m^2 ).
\end{eqnarray}
To renormalize the real part,
we need the subtracted dispersion relation appearing in Eq.~(\ref{eq:app:SDR})
for $n=2$,
\begin{eqnarray}
{\rm Re}\Sigma_1^F(P^2)
&=&
c_0 + c_1 P^2
+ \frac{P^4}\pi {\rm P} \int_{ m^2 }^{\infty} ds 
\frac{ {\rm Im}\Sigma_1^F(s) }{ s^2 ( s-P^2 ) }
\nonumber \\
&=&
c_0 + c_1 P^2  \notag \\
& & + \frac{ g^2 }{ 32\pi^2 } \left\{
\frac{ P^2 + 2m^2 }{ m^2 } \frac{ ( P^2-m^2 )^2 }{ P^4 }
\log \left| \frac{ P^2 - m^2 }{ m^2 } \right|
- \frac56 \frac{ P^2 }{ m^2 } - 2 + 2 \frac{ m^2 }{ P^2 }
\right\}, \notag \\
\label{eq:app:T=0vectorR}
\end{eqnarray}
where we have set $\alpha=0$.
We finds that the on-shell renormalization condition $\Sigma_1^F(P^2=0)=0$
is not sufficient to remove all the divergent parts.
Here we impose an additional condition, $c_1=0$,
to obtain the renormalized self-energy,
\begin{eqnarray}
{\rm Re}\Sigma_1^F(P^2)
&=&
\frac{ g^2 }{ 32\pi^2 } \left\{
\frac{ P^2 + 2m^2 }{ m^2 } \frac{ ( P^2-m^2 )^2 }{ P^4 }
\log \left| \frac{ P^2 - m^2 }{ m^2 } \right|
- \frac56 \frac{ P^2 }{ m^2 } - 2 + 2 \frac{ m^2 }{ P^2 }
\right\}. \notag \\
\end{eqnarray}
Therefore, the renormalized retarded self-energy is given by
\begin{eqnarray}
\Sigma_1^R(P^2)
&=&
\frac{ g^2 }{ 32\pi^2 } \left\{
\frac{ P^2 + 2m^2 }{ m^2 } \frac{ ( P^2-m^2 )^2 }{ P^4 }
\log \left| \frac{ P^2 - m^2 }{ m^2 } \right|
- \frac56 \frac{ P^2 }{ m^2 } - 2 + 2 \frac{ m^2 }{ P^2 }
\right\}.
\nonumber \\
&& -i \frac{ g^2 }{ 32\pi } 
\frac{ P^2 + 2m^2 }{ m^2 }
\frac{ ( P^2-m^2 )^2 }{ P^4 }
{\rm sgn}(\omega)\theta ( P^2 - m^2 ).
\label{eq:app:T=0vector}
\end{eqnarray}


\section{
Limiting Behavior of $\Sigma^R( \bm{p},\omega )$ in the Yukawa Model \\
with Vector Coupling}
\label{app:vector_asym}

In \S\ref{sec:vector}, we showed that the thermal mass
of the quark coupled with the vector boson in the Proca formalism
approaches $\omega=\sqrt{6}m$ in the high $T$ limit, 
where $m$ is the mass of the vector boson.
In this section, we prove this limiting behavior.
Because we are interested in the thermal mass,
we only consider the quark propagator at vanishing momentum
and omit the subscript and the argument of the momentum in the
self-energy, writing
$\Sigma_\pm(\bm{p}=0,\omega)$ as  $\Sigma(\omega)$,
in the following.

Let us first derive the analytic form of ${\rm Re}\Sigma(\omega)$
in the high $T$ limit, where
the $T$-independent part is negligible. 
Thus, we consider the $T$-dependent part ${\rm Re}\Sigma(\omega)_{T\neq0}$, 
which is 
calculated from
the unsubtracted dispersion relation in Eq.~(\ref{eq:Kramers-Kronig}),
\begin{eqnarray}
{\rm Re}\Sigma(\omega)_{T\ne0}
  = 
  -\frac1\pi \textrm{P}\int_{-\infty}^{\infty} dz
  \frac{ {\rm Im}\Sigma(z)_{T\ne0} }{ \omega - z }
  =
  -\frac{2\omega}\pi \textrm{P}\int_0^{\infty} dz
  \frac{ {\rm Im}\Sigma(z)_{T\ne0} }{ \omega^2 - z^2 }.
  \label{eq:Kramers-KronigApp}
\end{eqnarray}
In the second equality, we have used the relation
${\rm Im}\Sigma(\omega) = {\rm Im}\Sigma(-\omega)$,
which is valid for a massless quark
of vanishing momentum.
Here, ${\rm Im}\Sigma(\omega)_{T\ne0}$ is given by
\begin{eqnarray}
  {\rm Im}\Sigma(\omega)_{T\ne0}
  &=& -\frac{g^2}{64\pi} \frac{ \omega^2 + 2m^2 }{m^2}
  \frac{ ( \omega^2-m^2 )^2 }{ \omega^3 } \notag \\
  & & \times\left( \coth \frac{ \omega^2 + m^2 }{ 4T\omega }
+ \tanh \frac{ \omega^2 - m^2 }{ 4T\omega } 
- 2\theta( \omega^2-m^2 ) \right).
\label{eq:ImSigmaApp}
\end{eqnarray}
Substituting Eq.~(\ref{eq:ImSigmaApp}) into Eq.~(\ref{eq:Kramers-KronigApp}),
we have the following:
\begin{eqnarray}
{\rm Re}\Sigma(\omega)_{T\neq0}
&=& \frac{ \omega g^2 }{ 32\pi^2 } ( I_1(\omega) + I_2(\omega) ),
\label{eq:Re=I1+I2}
\end{eqnarray}
\begin{eqnarray}
I_1(\omega) 
&\equiv&
\textrm{P} \int_0^m dz \frac{1}{ \omega^2 - z^2 }
\frac{ z^2 + 2m^2 }{m^2}
\frac{ ( z^2-m^2 )^2 }{ z^3 }
\left[ \coth \frac{ z^2 + m^2 }{ 4Tz }
 - \tanh \frac{ z^2 - m^2 }{ 4Tz } \right] \notag \\
&=& \left(\frac{T}{\omega}\right)^2 \int_0^{1/\alpha}
    \frac{dx}{1-(\alpha \beta x)^2} \frac{(2+\alpha^2 x^2)
    (1-\alpha^2 x^2)^2}{x^3} \notag \\
& & \qquad \times \bigg[ \coth\bigg( \frac{1}{4x}+\frac{\alpha^2 x}{4}\bigg)
    -\tanh\bigg( \frac{1}{4x}-\frac{\alpha^2 x}{4}\bigg) \bigg],
\label{eq:I_1}
\\
I_2(\omega) 
&\equiv& 
\textrm{P} \int_m^\infty dz \frac{1}{ \omega^2 - z^2 }
\frac{ z^2 + 2m^2 }{m^2}
\frac{ ( z^2-m^2 )^2 }{ z^3 }
\left[ \coth \frac{ z^2 + m^2 }{ 4Tz }
 - \tanh \frac{ z^2 - m^2 }{ 4Tz } -2 \right] \notag \\
&=& -\left(\frac{T}{m}\right)^2 \int_0^{1/\alpha}
    \frac{dy}{1-\alpha^2 y^2/\beta^2}
    \frac{(1+2\alpha^2 y^2)(1-\alpha^2 y^2)^2}{y^3} \notag \\
& & \qquad \times \bigg[ \coth\bigg( \frac{1}{4y}+\frac{\alpha^2 y}{4}\bigg)
    +\tanh\bigg( \frac{1}{4y}-\frac{\alpha^2 y}{4}\bigg) -2\bigg],
\label{eq:I_2}
\end{eqnarray}
with $\alpha=m/T, \beta=m/\omega, x=z/(m\alpha)$ and
$y=m/(\alpha z)$.

In the high $T$ limit ($\alpha\to0$), the asymptotic forms of
$I_1$ and $I_2$ are given by
\begin{eqnarray}
  I_1(\omega) &\to& 2\left(\frac{T}{\omega}\right)^2 \int_0^{\infty}
  \frac{dx}{x^3} \bigg[ \coth\bigg( \frac{1}{4x} \bigg)
    -\tanh\bigg( \frac{1}{4x} \bigg) \bigg] \notag \\
  &=& 4\left(\frac{T}{\omega}\right)^2 \int_0^{\infty}
  dx x \bigg[ \frac{1}{\exp(x/2)-1} + \frac{1}{\exp(x/2)+1} \bigg] \notag\\
  &=& \frac{4\pi^2 T^2}{\omega^2}, \\
  I_2(\omega) &\to& -\left(\frac{T}{m}\right)^2 
  \int_0^{\infty} \frac{dy}{y^3} 
  \bigg[ \coth\bigg( \frac{1}{4y} \bigg)
    +\tanh\bigg( \frac{1}{4y}\bigg) -2\bigg] \notag \\
  &=& -2\left(\frac{T}{m}\right)^2  \int_0^{\infty} dy y
  \bigg[ \frac{1}{\exp(y/2)-1} - \frac{1}{\exp(y/2)+1} \bigg] \notag\\
  &=& -\frac{2\pi^2 T^2}{3m^2}.
\end{eqnarray}
We have confirmed that the correction terms due to the finite 
value of $\alpha$ are $O(\alpha \ln (1/\alpha))$.
Therefore, the self-energy in the high $T$ limit is given by
\begin{equation}
  \lim_{T\to0} {\rm Re}\Sigma(\omega) = \frac{g^2 T^2}{8}
  \left( \frac{1}{\omega}-\frac{\omega}{6m^2}\right).
\end{equation}
In this case, the quasi-dispersion relation is given by the solution of the
equation $\omega-{\rm Re}\Sigma(\omega)\approx -{\rm Re}\Sigma(\omega) =0$,
i.e., the zero of ${\rm Re}\Sigma(\omega)$.
Therefore, in this model,
the energy $\omega=\sqrt{6}m$ gives the thermal mass in the 
high $T$ limit.

\end{document}